\documentclass[11pt]{article}
 \usepackage{graphicx}  
\usepackage{amsmath}   
\usepackage[compress]{cite}
\usepackage{amssymb}   
\usepackage{bm} 
\usepackage{dcolumn}
\usepackage{ulem}
\usepackage{color}
\usepackage{mathtools}
\usepackage{mathrsfs}
\usepackage{amsfonts}
\usepackage{varioref}
\usepackage{textcomp}
\usepackage{geometry}
\usepackage{tcolorbox}
\usepackage{empheq}
\usepackage[active]{srcltx}
\usepackage{tikz}
\usepackage{lmodern}
\usepackage{braket}
\usetikzlibrary{decorations.pathmorphing}
\tikzset{snake it/.style={decorate, decoration=snake}}
\usepackage{fancybox}
\RequirePackage[colorlinks,citecolor=blue,urlcolor=magenta,linkcolor=blue]{hyperref}
\allowdisplaybreaks
 
\def\be{\begin{equation}}
\def\bea{\begin{eqnarray}}
\def\ee{\end{equation}}
\def\eea{\end{eqnarray}}

 \labelformat{equation}{Eq.\,(#1)} 
\labelformat{figure}{Fig.~#1} 
\labelformat{table}{Tab.~#1} 
 \title{\bf Supertranslations at Timelike Infinity}

\begin{document}

\title{\bf Supertranslations at Timelike Infinity }

\author{Sumanta Chakraborty$^{1,2}$, 
Debodirna Ghosh$^{3}$,
Sk Jahanur Hoque$^{4}$, \\ Aniket Khairnar$^{5}$ and Amitabh Virmani$^{3}$
\vspace{0.6cm} \\
$^{1}${\small{School of Mathematical and Computational Sciences}}
\\
{\small{Indian Association for Cultivation of Science, Kolkata 70032, India}}
\vspace{0.3cm} \\
$^{2}${\small{School of Physical Sciences}}
\\
{\small{Indian Association for Cultivation of Science, Kolkata 70032, India}}
\vspace{0.3cm} \\
$^{3}${\small{Chennai Mathematical Institute,}} 
\\ {\small{H1 SIPCOT IT Park, Kelambakkam, Tamil Nadu 603103, India }}
\vspace{0.3cm} \\
$^{4}${\small{Institute of Theoretical Physics,
Faculty of Mathematics and Physics, Charles University,}}
\\
{\small{V~Hole\v{s}ovi\v{c}k\'ach 2, 180~00 Prague 8, Czech Republic}}
\vspace{0.3cm} \\
$^{5}${\small{Department of Physics and Astronomy,
The University of Mississippi}}
\\
{\small{P.O. Box 1848, University, MS 38677 USA}}
}
\maketitle

\begin{abstract}
We propose a definition of asymptotic flatness at timelike infinity in four spacetime dimensions. We present a detailed study of the asymptotic equations of motion and the action of supertranslations on asymptotic fields. We show that the  Lee-Wald symplectic form $\Omega (g, \delta_1 g, \delta_2 g)$ does not get contributions from future timelike infinity with our boundary conditions.  As a result, the ``future charges'' can be computed on any two-dimensional surface surrounding the sources at timelike infinity. We present expressions for  supertranslation and Lorentz charges. 
\end{abstract}
\newpage
\tableofcontents
\newpage

\section{Introduction}
The asymptotic properties of gravity have been studied for decades \cite{Bondi:1962px, Sachs:1962wk, Strominger:2013jfa, He:2014laa} in the context of asymptotically flat spacetimes at null infinity, see \cite{Ashtekar:2014zsa, Alessio:2017lps, Madler:2016xju, Strominger:2017zoo, Ashtekar:2018lor, Aneesh:2021uzk} for recent reviews. Much of these studies are rightly motivated by the need to understand the intricate nature of gravitational radiation. One remarkable outcome of these studies was the discovery of the infinite-dimensional Bondi-Metzner-Sachs (BMS) group of asymptotic symmetries in asymptotically flat spacetimes at null infinity. Recent works have shown that the BMS group is related to the infrared properties of gravity, namely soft-theorems and memory effects~\cite{Strominger:2017zoo}. See references \cite{Ashtekar:1981bq, Ashtekar:1981sf, Ashtekar:1987tt} for earlier works on these issues and  the reviews~\cite{Strominger:2017zoo, Ashtekar:2018lor}  for further references. The exploration of the connections between  the BMS group, soft-theorems, and memory effects
have led to enormous activity. Further enlargements of the BMS group  \cite{Barnich:2010eb, Barnich:2010ojg, Campiglia:2014yka, Freidel:2021fxf, Gupta:2021cwo, Aneesh:2021uzk}  have also been argued to be relevant.

The BMS symmetries are exact symmetries of General Relativity, in the sense that they leave the action invariant up to a surface term. This strongly suggests that they should be visible in any description, in particular at spatial infinity and timelike infinity, provided boundary conditions at spatial and timelike infinity are compatible with boundary conditions at null infinity. This poses the dynamical question: how to relate boundary conditions at null, spatial, and timelike infinity?  The answer to this question remains poorly understood, and therein lies the key to many unresolved issues. The importance of understanding these issues has been stressed by Friedrich in a recent article~\cite{Friedrich:2017cjg}. 

On the specific question of  BMS symmetries at spatial infinity there has been a lot of progress in recent years, motivated in part by the need to understand the relation between  the BMS group, soft-theorems, and memory effects.  Earlier investigations of the asymptotic symmetries at spatial infinity \cite{Arnowitt:1962hi, Regge:1974zd, Ashtekar:1978zz, Ashtekar:1991vb} successfully found boundary conditions that gave Poincar\'e group as asymptotic symmetries. 

This situation was exhilarating on one hand and disappointing on the  other. Exhilarating because at least at spatial infinity there are consistent boundary conditions that  lead to Poincar\'e group as asymptotic symmetries whereas at null-infinity this does not seem desirable. Disappointing because the lack of understanding of the BMS symmetries at spatial infinity means that the relation between boundary conditions at null and spatial infinity is incomplete. This had remained a deep puzzle for many years.

Henneaux and Troessaert \cite{Troessaert:2017jcm, Henneaux:2018cst, Henneaux:2018hdj} in a series of paper have resolved this tension, both in the cylindrical representation and the hyperbolic representation of spatial infinity. They have proposed boundary conditions at spatial infinity that are invariant under BMS symmetries. The BMS symmetries have non-trivial action on the fields and have generically non-zero charges. They have also related BMS generators at spatial infinity to BMS generators at past and future null infinity. Other works in this direction include~\cite{Prabhu:2019fsp, Prabhu:2019daz}.

The situation at timelike infinity remains much less developed. Following works at spatial infinity~\cite{Ashtekar:1978zz, Ashtekar:1991vb}, earlier work \cite{cutler, porrill, Gen:1997az} had proposed boundary conditions 
that gave Poincar\'e group as asymptotic symmetries. 
To the best of our knowledge, 
 no attempt has been made to realise 
BMS symmetries at timelike infinity in the non-linear theory.  
Motivated by the relation between the BMS group and soft-theorems,  these issues were addressed in the linearised gravity in \cite{Campiglia:2015kxa, H:2020eei}, though the main focus in these papers is somewhat different.
 The main aim of this paper is to present boundary conditions in non-linear general relativity at timelike infinity that realise BMS symmetries in the sense that  BMS symmetries have non-trivial action on the fields and have generically non-zero charges. Our work in motivated by the corresponding developments at spatial infinity
  \cite{Troessaert:2017jcm, Henneaux:2018cst, Henneaux:2018hdj, Prabhu:2019fsp, Prabhu:2019daz}.

Such a study is important for several reasons.    Over the last two decades, it has been argued in a variety of contexts  that stationary black holes also possess an infinite number of symmetries in the near horizon region \cite{Koga:2001vq, Donnay:2015abr,  Hawking:2016msc, Hawking:2016sgy, Carlip:2017xne, Chandrasekaran:2018aop}.\footnote{The symmetry groups in these papers do not coincide. This is so because different authors preserve different structures: some prefer to preserve a particular geometric structure on the null surface, whereas others preserve the near horizon geometry.} 
 Typically a class of these symmetries is similar to supertranslations at null infinity.  It is believed that global charges associated with supertranslations receive contributions from the  horizon as well as from null infinity.  Clearly, a complete discussion of conservation laws associated with supertranslations requires a detailed understanding of how the symmetries at the horizon relate to the symmetries at null infinity. However, this has not been understood.\footnote{To some extent these issues were explored in \cite{Hawking:2016msc, Hawking:2016sgy}. In these references, advanced Bondi coordinates are used; however, since advanced Bondi coordinates do not cover future null infinity, the relation between symmetries at future null infinity and the future horizon remains unexplored. These points were recently  emphasised in \cite{Fernandes:2020jto, Donnay:2020fof}. More broadly, in recent years several studies of null boundaries have advanced our knowledge of fluxes along null surfaces \cite{Parattu:2015gga, Parattu:2016trq, Hopfmuller:2016scf, Grumiller:2020vvv, Adami:2021nnf}.} Toy model studies include \cite{Fernandes:2020jto, Donnay:2020fof}. It has been suggested by several authors, specifically  by Chandrasekaran, Flanagan, and Prabhu in \cite{Chandrasekaran:2018aop} that timelike infinity can be used to  relate symmetries at the horizon to symmetries at null infinity.

In this paper we only focus on timelike infinity with perhaps the simplest boundary conditions that allow for 
the BMS symmetries. The dynamical questions on the relationship of our boundary conditions to null and spatial infinity is beyond the scope of this work. Issues related to further enlargement of BMS symmetries \cite{Barnich:2010eb, Barnich:2010ojg, Campiglia:2014yka, Gupta:2021cwo} are also beyond the scope of this work. We hope to return to these questions in future works.

The rest of the paper is organised as follows. In section \ref{sec:asym_flatness}, we introduce our notion of asymptotic flatness at timelike infinity. Many of the calculations here are direct translations of those at spatial infinity. Having said so, we must add that the literature at spatial infinity is fairly large and confusing. Therefore, it is absolutely essential to work-out things from the start to the end for timelike infinity separately.  In section \ref{sec:EOM}, a detailed study of the asymptotic equations of motion is presented. In section \ref{sec:charges}, expressions for supertranslation and Lorentz charges are proposed. In section  \ref{sec:schw}, the Schwarzschild solution is written in a form such that it manifestly satisfies our boundary conditions. In section \ref{sec:horizon}, some general remarks on supertranslations are made. We close with a brief discussion in section \ref{sec:disc}. Dynamical questions regarding the non-triviality of our construction, i.e., whether non-trivial radiative spacetimes exist that satisfy our boundary conditions at timelike infinity requires a separate investigation.

\section{Asymptotic flatness at timelike infinity}
\label{sec:asym_flatness}
In this section we introduce our notion of asymptotic flatness at timelike infinity. 
It is based on the corresponding notion introduced by Beig and Schmidt \cite{BeigSchmidt, Beig:1983sw} at spatial infinity, which has been extensively studied over the years \cite{Ashtekar:1990gc, Ashtekar:2008jw, Compere:2011db, Compere:2011ve, Virmani:2011aa}. We work with a coordinate based definition. If needed, our results can be readily translated to geometric frameworks. 
A notion of 
asymptotic flatness at timelike infinity in the geometrical framework of Ashtekar-Hansen~\cite{Ashtekar:1978zz} was introduced by Cutler \cite{cutler} and Porrill \cite{porrill}. A closely related notion in the geometrical framework of Ashtekar-Romano~\cite{Ashtekar:1991vb} was discussed by  Gen and Shiromizu \cite{Gen:1997az}.  Our notion  is different from all these previous works, as we allow a class of spi-supertranslations to act as asymptotic symmetries  at timelike infinity.

\subsection{Asymptotic metric}

To introduce our notion of asymptotic flatness at timelike infinity we start by introducing a set of ``polar coordinates'' for Minkowski spacetime $\{\tau, \rho, \theta, \varphi\}$ as follows
\begin{align}
&\eta_{\mu \nu } x^\mu x^\nu = - \tau^2, &
&\frac{r}{t} = \frac{\rho}{\sqrt{1 + \rho^2}},
\end{align}
where $\eta_{\mu \nu} = \verb+diag+\{-1,1,1,1\}$ and $x^\mu$ are a standard set of cartesian coordinates and $r^2 = (x^1)^2 + (x^2)^2 + (x^3)^2$. In these coordinates flat spacetime metric takes the form
\bea
ds^2 &=& - d\tau^2 + \tau^2 \left( \frac{d\rho^2}{1+ \rho^2} + \rho^2 (d\theta^2 + \sin^2 \theta d \varphi^2) \right),\\
&\equiv &  - d\tau^2 + \tau^2 h^{(0)}_{ab} d\phi^a d\phi^b
\eea
where we denote coordinates $\{\rho, \theta, \varphi\}$ collectively as $\phi^a$. Metric $h^{(0)}_{ab}$ is the unit metric on  Euclidean AdS$_3$ hyperboloid ${\cal H}$.

We start by considering a general class of  spacetime admitting an expansion at timelike infinity of the form
\begin{align}\label{asymp_metric_01}
g_{\mu \nu}=\eta_{\mu \nu}+\sum_{n=1}^{m}\frac{\ell_{\mu \nu}^{(n)} }{\tau^{n}}+ o(\tau^{-m}),
\end{align} 
where $\ell_{\mu \nu}^{(n)}$, for each $n$,  is a function of $\frac{x^{\sigma}}{\tau}$.
Following Beig and Schmidt~\cite{Beig:1983sw}, this metric can be put in the following more convenient form
\begin{align}\label{Beig_Schmidt_form}
&ds^{2}=-N^{2}d\tau^{2}+h_{ab}d\phi^{a}d\phi^{b}, 
\end{align}
where
\begin{align}
&N=1+\frac{\sigma(\phi^{a})}{\tau}, \label{Beig_Schmidt_form1} \\
&h_{ab}=\tau^{2}\left[h^{(0)}_{ab}(\phi^{c})+\frac{1}{\tau}h^{(1)}_{ab}(\phi^{c})+\frac{1}{\tau^{2}}h^{(2)}_{ab}(\phi^{c})+\mathcal{O}\left(\frac{1}{\tau^{3}}\right)\right].\label{Beig_Schmidt_form2}
\end{align}
A derivation of the form of the metric \ref{Beig_Schmidt_form}--\ref{Beig_Schmidt_form2} starting from \ref{asymp_metric_01} is given in appendix \ref{asymptotic_metric}. We define asymptotically flat spacetimes at timelike infinity as spacetimes admitting an asymptotic expansion as in \ref{Beig_Schmidt_form}--\ref{Beig_Schmidt_form2}. Further boundary conditions will be specified below. We will comment on the smoothness of fields $\sigma$, $h_{ab}^{(1)}$, $h_{ab}^{(2)}$, etc. on EAdS$_3$ hyperboloid $\cal{H}$ at a later stage.

 \subsection{Supertranslation at timelike infinity}
A natural question to ask is what is the set of  diffeomorphisms preserving the form of the metric \ref{Beig_Schmidt_form}--\ref{Beig_Schmidt_form2}.\footnote{In the context of spatial infinity this question has been analysed in great detail by many authors over the years; see \cite{Ashtekar:2008jw, Compere:2011db} for a concise summary of the earlier results. }
In particular, if supertranslations are genuine symmetries of general relativity then they should also be realisable at timelike infinity. 
In order to spell out our boundary conditions explicitly, we start by looking at the action of supertranslations on asymptotic fields.

\subsubsection{First order}
As shown in detail in appendix \ref{supertranslations_app}, the diffeomorphism 
\begin{align}
\label{first_order_super}
&\tau=\bar{\tau}-\omega(\bar{\phi}^{a})+\mathcal{O}\left(\frac{1}{\bar{\tau}}\right), \\
&\phi^{a}=\bar{\phi}^{a}+\frac{1}{\bar{\tau}}h^{(0)ab}\partial_{b}\omega(\bar{\phi}^{c})+\mathcal{O}\left(\frac{1}{\bar{\tau}^{2}}\right),
\label{first_order_super2}
\end{align}
preserves the asymptotic form of the metric to order $\frac{1}{\tau}$.  Here $\omega (\phi^a)$ is an arbitrary function on $\cal{H}$ that determines the higher order terms in the diffeomorphism uniquely. When $\omega(\phi^a)$ is in the four-parameter class of solutions of 
\be
\mathcal{D}_{a} \mathcal{D}_{b}\,\omega -\omega h_{ab}^{(0)} =0, \label{translations}
\ee
with $\mathcal{D}_{a}$ being the covariant derivative on  $\cal{H}$ compatible with metric $h_{ab}^{(0)}$, the transformation \ref{first_order_super}--\ref{first_order_super2} correspond to a translation. More generally, the above diffeomorphism corresponds to  a supertranslation.  The four functions satisfying 
\eqref{translations} are \be \left\{ \sqrt{1 +\rho^2}, \rho \cos \theta, \rho \sin \theta \sin \phi,  \rho \sin \theta \cos \phi  \right \}, \ee
representing respectively, the time-translation and three-spatial translations.

Under general supertranslation \ref{first_order_super}--\ref{first_order_super2}, the zeroth order field $h^{(0)}_{ab}$ does not transform,
\be
h^{(0)}_{ab} \rightarrow h^{(0)}_{ab}, 
\ee
whereas the first order fields transform as,
 \begin{align}
&\sigma \rightarrow  \sigma, \\
& h_{ab}^{(1)} \rightarrow h_{ab}^{(1)} + 2\mathcal{D}_{a} \mathcal{D}_{b}\,\omega - 2\omega h_{ab}^{(0)}. \label{change_first_order}
\end{align}
We define, 
\be
k_{ab} : =h_{ab}^{(1)}+2\sigma h^{(0)}_{ab}. \label{def_kab}
\ee
 It follows from \ref{change_first_order} that under  general supertranslation,
\be
k_{ab} \rightarrow k_{ab} + 2\mathcal{D}_{a}\mathcal{D}_{b}\omega - 2\omega h_{ab}^{(0)}.
\label{change_first_order_k}
\ee
Now, there are two natural set of boundary conditions to consider. First, one can dispose of all the supertranslations by demanding
\be
k_{ab} = 0.
\ee
These are the boundary conditions used in \cite{cutler, porrill, Gen:1997az}. As is clear from \ref{change_first_order_k} that with these boundary conditions, supertranslations are not allowed asymptotic symmetries.  In the class of diffeomorphisms \ref{first_order_super}--\ref{first_order_super2} only translations (cf.~\ref{translations}) are allowed asymptotic symmetries.

Second, motivated by the work on spatial infinity \cite{Compere:2011ve} and \cite{Troessaert:2017jcm, Henneaux:2018cst} one can choose,
\be
k := h^{(0)ab} k_{ab}= 0. \label{BC_k_0}
\ee
The requirement that the trace of $k_{ab}$ vanishes should be invariant under supertranslations. From \ref{change_first_order_k} we therefore deduce that the following differential equation for the function $\omega$,
\begin{align}
\left(\square-3\right)\omega=0.
\end{align}
This is the class of supertranslations we work with in this paper. Here $\square$ is the Laplacian on $\cal{H}$: $\square = {\cal D}_a {\cal D}^a$.

There can be other classes of transformations with appropriately modified notions of asymptotic flatness, e.g., logarithmic translations, superrotations, more general supertranslations, that one can explore. We do not study them in this work. Very likely, our considerations can be extended to include a study of logarithmic translations following, say, \cite{Compere:2011ve}. However, how superrotations at timelike infinity \cite{Campiglia:2015kxa, H:2020eei, Campiglia:2015lxa} can feature in such an analysis is not clear to us. Naively, the introduction of 
superrotations does not look compatible with the zeroth order equations of motion in the $1/\tau$ expansion.\footnote{We thank the anonymous referee for suggesting us to add these comments.}

 
\subsubsection{Second order}
It is of interest to study the action of supertranslations on the second order fields.  At second order the diffeomorphism presented in \ref{first_order_super}--\ref{first_order_super2} generalises to,
\begin{align}
&\quad \tau=\bar{\tau}-\omega(\bar{\phi}^{a})+\frac{1}{\bar{\tau}}F^{(2)}(\bar{\phi}^{a})+\mathcal{O}\left(\frac{1}{\bar{\tau}^{2}}\right)~,
\\
&\quad \phi^{a}=\bar{\phi}^{a}+\frac{1}{\bar{\tau}}h^{(0)ab}\partial_{b}\omega(\bar{\phi}^{c})+\frac{1}{\bar{\tau}^{2}}G^{(2)a}(\bar{\phi}^{c})+\mathcal{O}\left(\frac{1}{\bar{\tau}^{3}}\right)~,
\end{align}
where the functions $F^{(2)}(\bar{\phi}^{a})$ and $G^{(2)a}(\bar{\phi}^{c})$ are uniquely fixed in terms of the function $\omega(\bar{\phi}^{a})$ by the requirement that the form of the metric should remain the same to order $\frac{1}{\tau^2}$.
We apply the above transformations and expand the metric in \ref{Beig_Schmidt_form}--\ref{Beig_Schmidt_form2} upto second order. Using the boundary condition $k=0$, we find, 
\begin{align}
\label{change_in_Secondh}
h^{(2)}_{ab} \rightarrow &~ h^{(2)}_{ab}-\omega k_{ab}+\omega^{c} \mathcal{D}_{c}\,k_{ab}+\frac{1}{2}\omega^{c}_{b}\,k_{ac}+\frac{1}{2}\omega^{c}_{a}\,k_{bc}-\frac{1}{2}\omega^{c}\left(\mathcal{D}_{a}\,k_{bc}\right)-\frac{1}{2}\omega^{c}\left(\mathcal{D}_{b}\,k_{ac}\right)
\nonumber
\\
& +2\sigma \omega h^{(0)}_{ab}+\sigma_{(a}\omega_{b)}-\sigma \omega_{ab}-\sigma_{c(a}\omega^{c}_{b)}-\sigma_{c}\omega^{c}{}_{(ab)}+\left(\sigma \leftrightarrow \omega \right)
\nonumber
\\
& +\omega^{2}h^{(0)}_{ab}-2\omega \omega_{ab}+\omega^{c}_{a}\omega_{bc}~.
\end{align}
Here $\sigma_a = \mathcal{D}_{a} \sigma, \mathcal{D}_{a} \mathcal{D}_{b} \sigma = \sigma_{ba}, \mathcal{D}_{a} \mathcal{D}_{b} \mathcal{D}_{c} \sigma = \sigma_{cba}$ etc.~and similarly for $\omega$.
A detailed derivation is given in appendix \ref{supertranslations_app}.

 A non-trivial consistency check on this expression is presented in appendix \ref{a_consistency_check}. There we consider doing  a supertranslation on flat spacetime. We begin with (for flat spacetime)
\be
\sigma = 0, \quad h^{(1)}_{ab} =0, \quad h^{(2)}_{ab} =0.
\ee
We note that $\sigma$ does not change under supertranslations. Thus for the supertranslated spacetime too  $\sigma = 0$. From \ref{change_first_order}, it follows that for the supertranslated spacetime
\be
h^{(1)}_{ab}= k_{ab} =-2\omega h^{(0)}_{ab}+2\omega_{ab}, \label{h1_flat_omegaMT}
\ee
and from  \ref{change_in_Secondh}, it follows that 
\begin{align}
h^{(2)}_{ab}
=  \omega^{2}h^{(0)}_{ab}-2\omega \omega_{ab}+\omega^{c}_{a}\omega_{bc}. \label{h2_flat_omegaMT}
\end{align}
In appendix  \ref{a_consistency_check}, we check that the expression in \ref{h1_flat_omegaMT} for $k_{ab}$ and \ref{h2_flat_omegaMT} for $h^{(2)}_{ab}$  are consistent with  the second order equations of motion.\footnote{An expression for corresponding transformation of $h^{(2)}_{ab} $ at spatial infinity was reported in equation (4.111) of \cite{Compere:2011ve}. All the $\omega$-$\omega$-terms, the analog of the third line in \ref{change_in_Secondh}, are missing there.   We note that the action of supertranslations at the second order has not been much discussed in the literature;  comments appear in \cite{Compere:2011ve, Virmani:2011aa}, though neither of these papers present any details on this specific calculation. We hope that the reader will find our appendices \ref{supertranslations_app} and \ref{a_consistency_check} useful. The action of \textit{translations} was discussed in \cite{Beig:1983sw}.}

\section{Asymptotic expansion of the equation of motion}

\label{sec:EOM}

Einstein's equations can be split into 3+1 form, providing a set of three equations appropriately projected along normal direction to constant $\tau$ hypersurface. The split provides the Hamiltonian and momentum constraints, and the    evolution equation for the metric on the 3-dimensional $\tau= \textrm{constant} $ hypersurface. These equations read,
\begin{align}
& H \equiv \frac{1}{N}\partial_{\tau}K+K_{ab}K^{ab}-\frac{1}{N}h^{ab}D_{a}D_{b}N =0,
\\
& H_{a} \equiv D_{b}K^{b}_{a}-D_{a}K =0, \\
& H_{ab} \equiv \mathcal{R}_{ab}+\frac{1}{N}\partial_{\tau}K_{ab}-2K_{ac}K^{c}_{b}+KK_{ab}-\frac{1}{N}D_{a}D_{b}N =0.
\end{align}
Here $D$ is the covariant derivative  compatible with metric $h_{ab}$ and $K_{ab} = \frac{1}{2 N} \partial_\tau h_{ab}$ is the extrinsic curvature of the constant $\tau$ hypersurface.

These equations can be expanded in inverse powers of $\tau$  as,
\begin{align}
&\quad H\equiv \frac{H^{(0)}}{\tau^{2}}+\frac{H^{(1)}}{\tau^{3}}+\frac{H^{(2)}}{\tau^{4}}+\mathcal{O}\left(\frac{1}{\tau^{5}}\right),
\\
&\quad H_{a} \equiv \frac{H_{a}^{(0)}}{\tau}+\frac{H_{a}^{(1)}}{\tau^{2}}+\frac{H_{a}^{(2)}}{\tau^{3}} + \mathcal{O}\left(\frac{1}{\tau^{4}}\right),
\\
&\quad H_{ab} \equiv H^{(0)}_{ab}+\frac{1}{\tau}H^{(1)}_{ab}+\frac{1}{\tau^{2}}H^{(2)}_{ab} + \mathcal{O}\left(\frac{1}{\tau^{3}}\right).
\end{align}
The expansion coefficients at zeroth, first, and second order are summarised in the following subsections. A detailed derivation of these results is given in appendix \ref{appendix_EOM}.

\subsection{Zeroth and first order}
At zeroth order, the Hamiltonian and the momentum constraints are identically satisfied. 
The evolution equation implies that the three-dimensional metric $h^{(0)}_{ab}$ on ${\cal H}$  must satisfy,
\begin{align}
H_{ab}^{(0)}&=\mathcal{R}^{(0)}_{ab}+2h^{(0)}_{ab} =0.
\end{align}
This condition implies that ${\cal H}$ is maximally symmetric with $\mathcal{R}^{(0)}=-6$ and the Riemann tensor is given by, 
\begin{align}\label{zeroth_order_02}
\mathcal{R}^{(0)}_{abcd}=\frac{\mathcal{R}^{(0)}}{6}\left(h^{(0)}_{ac}h^{(0)}_{bd}-h^{(0)}_{ad}h^{(0)}_{bc}\right)
=-h^{(0)}_{ac}h^{(0)}_{bd}+h^{(0)}_{ad}h^{(0)}_{bc}~.
\end{align}
Thus ${\cal H}$ is Euclidean AdS$_3$ space, as noted earlier.

At first order, the Hamiltonian constraint gives,
\begin{align}
\label{first_order_01}
H^{(1)}&=\left(-\square+3\right)\sigma =0.
\end{align}
The momentum constraint gives,
\begin{align}
\label{first_order_02}
\mathcal{D}^{b} k_{ab}= \mathcal{D}_{a} k.
\end{align}
The evolution equations  $H^{(1)}_{ab}=0$ gives,
\begin{align}
\label{first_order_03}
\left(\square + 3 \right) k_{ab}= \mathcal{D}_{a} \mathcal{D}_{b} k + k h^{(0)}_{ab}.
\end{align}
Boundary conditions presented in \ref{BC_k_0} further simplify these equations to
\begin{align}
\mathcal{D}^{b} k_{ab} &= 0,  &
\left(\square + 3 \right) k_{ab}&= 0.
\end{align}

 \subsection{Second order}

At second order, the Hamiltonian constraint takes the form, 
\begin{align}\label{second_order_01}
h^{(2)}=12\sigma^{2}+\frac{1}{4}k^{ab}\,k_{ab}-k^{ab}\mathcal{D}_{a}\,\mathcal{D}_{b}\sigma- \mathcal{D}_{c}\sigma \,\mathcal{D}^{c}\sigma,
\end{align}
where $h^{(2)}$ is the trace of  $h^{(2)}_{ab}$, $h^{(2)} = h^{(0)ab} h^{(2)}_{ab}$. In arriving at this equation we have used the boundary condition $k=0$ cf.~\ref{BC_k_0} and the first order equations of motion.
The momentum constraint reads,
\begin{align}\label{second_order_02}
\mathcal{D}_{b}h^{(2)b}_{a}=\frac{1}{2}k^{bp}\left(\mathcal{D}_{b}\,k_{pa}\right)
+\mathcal{D}_{a}\left(-\frac{1}{8}\,k^{bc}\,k_{bc}+8\sigma^{2}-k^{ab}\mathcal{D}_{a}\,\mathcal{D}_{b}\sigma- \mathcal{D}_{c}\sigma \mathcal{D}^{c}\sigma\right).
\end{align}
The evolution equation $H^{(2)}_{ab}=0$ yields, 
\begin{align}\label{second_order_03}
\left(\square+2\right)h^{(2)}_{ab}=S^{(kk)}_{ab}+S^{(k\sigma)}_{ab}+S^{(\sigma \sigma)}_{ab},
\end{align}
where the non-linear source terms have the following expressions,
\begin{align}
S^{(kk)}_{ab}= & \left(\mathcal{D}_{c}\,k_{d(a}\,\mathcal{D}_{b)}\,k^{cd}\right)-\frac{1}{2}\mathcal{D}_{a}\,k^{cd}\,\mathcal{D}_{b}\,k_{cd}
+\left(\mathcal{D}^{c}\,k_{ad}\right)\left(\mathcal{D}_{c}k^{d}_{b}\right)-\left(\mathcal{D}^{c}\,k_{ad}\right)\left(\mathcal{D}^{d}\,k_{bc}\right)
\nonumber
\\
& -k^{p}_{a}\,k_{pb} + k^{cd}\left(\mathcal{D}_{c}\,\mathcal{D}_{d}\,k_{ab}- \mathcal{D}_{c}\mathcal{D}_{(a}\,k_{b)d}\right)~,
\\
S^{(k\sigma)}_{ab}= & - \mathcal{D}_{a}\mathcal{D}_{b}\left(k^{cd}\mathcal{D}_{c}\,\mathcal{D}_{d}\sigma\right)+4\mathcal{D}^{c}\sigma \left(-\mathcal{D}_{c}k_{ab}+ \mathcal{D}_{(a}\,k_{b)c}\right)-4\sigma k_{ab}
\nonumber
\\
& +\left(-2h^{(0)}_{ab}k^{cd}\mathcal{D}_{c}\,\mathcal{D}_{d}\sigma +4k^{cd}h^{(0)}_{d(a}\mathcal{D}_{b)}\mathcal{D}_{c}\sigma\right)~,
\\
S^{(\sigma \sigma)}_{ab} = & \mathcal{D}_{a}\,\mathcal{D}_{b}\left(5\sigma^{2}- \mathcal{D}_{c}\sigma \mathcal{D}^{c}\sigma\right)
+h^{(0)}_{ab}\left(18\sigma^{2}+4\mathcal{D}^{c}\sigma \mathcal{D}_{c}\sigma\right)
+4\sigma \mathcal{D}_{a}\mathcal{D}_{b}\sigma~.
\end{align}

The second order equations of motion, in the form presented above, with more restrictive boundary condition $k_{ab} = 0$ take a particularly nice form and can be concisely presented in terms of the electric and magnetic parts of the Weyl tensors, as is the case at spatial infinity \cite{Mann:2006bd, Mann:2008ay, Compere:2011db}. These results are presented in appendix \ref{Weyl_tensor_expansion}.

\section{Charges at timelike infinity}

\label{sec:charges}

Next we would like to understand contributions from timelike infinity to the Iyer-Wald global charges~\cite{Lee:1990nz, Iyer:1994ys} (see also \cite{Chandrasekaran:2018aop})  for supertranslations and Lorentz symmetries. To this end, we compute contributions from timelike infinity to the Lee-Wald symplectic form. This computation is presented in section \ref{sec:Lee_Wald}. We find that with our boundary conditions this contribution vanishes. It has been suggested by several authors\footnote{See for example section 7 of \cite{Chandrasekaran:2018aop}.} that this should be the case with appropriate boundary conditions at timelike infinity.  As a result, ``future charges'' can be computed on any two-dimensional topologically-spherical surface surrounding the ``sources'' at timelike infinity. We present charge expressions in section \ref{sec:charges_timelike}. Some further properties of these charges are studied in section \ref{sec:charges_commutator}.

What are these sources at timelike infinity?  Note that bound objects (and fields) reach timelike infinity. Fields close to these bound objects do not become weak and cannot be regarded as asymptotic fields in the usual sense.  In the $\tau \to \infty$ limit, it is convenient  to regard individual bound systems, gravitationally unbound relative to each other, as finite number of points on the timelike infinity hyperboloid $\mathcal{H}$. These points serve as sources for the charge integrals. This picture will become more clear in section \ref{sec:schw} where we discuss the Schwarzschild solution in the $\tau \to \infty$ limit.

\subsection{Contributions to the Lee-Wald symplectic form}
\label{sec:Lee_Wald}

\begin{figure}[t!]
\begin{center}
 \includegraphics[width=0.4\textwidth]{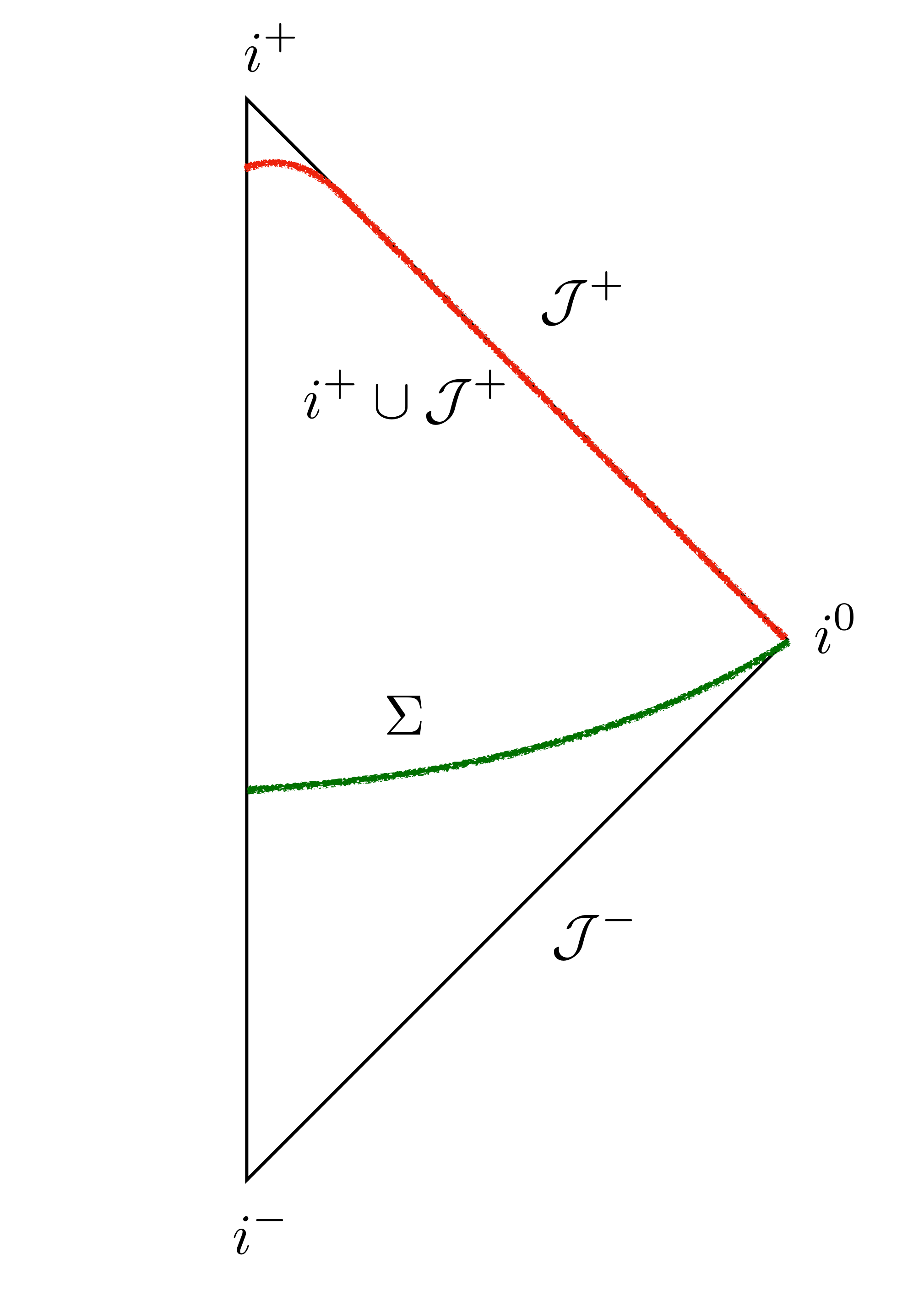}
 \caption{\sf Consider a spacetime with no horizons. The components of the boundary are $\mathcal{J}^-$, $\mathcal{J}^+$ and the points at infinity $i^-$, $i^0$, and $i^+$. Since the first variation of global charges is invariant under local deformations of the Cauchy surface $\Sigma$, one can deform $\Sigma$ in the far future to $i^+ \cup \mathcal{J}^+$. Then, the first variation of the Iyer-Wald global charges satisfies $\delta Q_\xi (\Sigma) = \delta Q_\xi ( \mathcal{J}^+) + \delta Q_\xi ( i^+)$. With our boundary conditions $\delta Q_\xi ( i^+) =0$. }
\label{fig:figure_flat}
\end{center}
\end{figure}

Consider a spacetime with no horizons. The components of the boundary are the past and future null infinity $\mathcal{J}^-$, $\mathcal{J}^+$ and the points (in the Penrose diagram in \ref{fig:figure_flat}) past and future timelike infinity $i^-$, $i^+$, and spatial infinity $i^0$. Since the global charge variation is invariant under local deformations of the Cauchy surface $\Sigma$, one can deform $\Sigma$ in the far future to $i^+ \cup \mathcal{J}^+$. Then, the first variation of the Iyer-Wald global charges satisfies 
\be
\delta Q_\xi (\Sigma) = \delta Q_\xi ( \mathcal{J}^+) + \delta Q_\xi ( i^+). \label{Iyer-Wald-null-timelike}
\ee
With our boundary conditions we now show that $\delta Q_\xi ( i^+) =0$. This is schematically shown in \ref{fig:figure_flat}. Recall that
\be
\delta Q_\xi (\Sigma) = \Omega(g,\delta g,\pounds_\xi g).
\ee

The computation proceeds as follows. The Lee-Wald symplectic form \cite{Lee:1990nz, Iyer:1994ys} is
\begin{align}
\Omega(g,\delta_{1}g,\delta_{2}g)=\int_{\Sigma}\boldsymbol{\omega}(g,\delta_{1}g,\delta_{2}g) = \int_{\Sigma} \omega^\gamma n_\gamma \, \sqrt{h} \,d^3 x,
\end{align}
where
\begin{align}
\omega^{\gamma}&=P^{\gamma \nu \alpha \beta \mu \delta} \left[\delta_{2}g_{\nu \alpha}\nabla_{\beta}\delta_{1}g_{\mu \delta}-\left(1\leftrightarrow 2\right) \right],
\nonumber
\\
P^{\gamma \nu \alpha \beta \mu \delta}&=g^{\gamma \mu}g^{\delta \nu}g^{\alpha \beta}-\frac{1}{2}g^{\gamma \beta}g^{\nu \mu}g^{\delta \alpha}
-\frac{1}{2}g^{\gamma \nu}g^{\alpha \beta}g^{\mu \delta}-\frac{1}{2}g^{\nu \alpha}g^{\gamma \mu}g^{\delta \beta}
+\frac{1}{2}g^{\nu \alpha}g^{\gamma \beta}g^{\mu \delta},
\end{align}
and where $n^\gamma$ is the unit normal to the hypersurface $\Sigma$, 
\be
n = - N d \tau,
\ee
and $\nabla_{\alpha}$ is the covariant derivative compatible with the spacetime metric $g_{\mu \nu}$. We choose the hypersurface $\Sigma$ to be a $\tau=\textrm{constant}$ surface. The volume factor $\sqrt{h} \, d^3 x$ grows as $\tau^3$  in the $\tau \to \infty$ limit. The aim, therefore, is to determine how $\omega^\gamma n_\gamma$ behaves in the $\tau \to \infty$ limit.  On $\tau=\textrm{constant}$ hypersurface,
\be
-\omega^{\gamma}n_{\gamma}=N\omega^{\tau}=\omega^{\tau} \left( 1 +\mathcal{O}(1/\tau) \right). 
\ee
As a result, the problem simply reduces to analysing the behaviour of $\omega^{\tau}$ in the $\tau \to \infty$ limit. For our purposes, the $\omega^\gamma$ expression can be written in a more convenient form as follows,
\begin{align}
\omega^{\gamma}&=g^{\gamma \mu}g^{\delta \nu}g^{\alpha \beta}\left(\delta_{2}g_{\nu \alpha}\nabla_{\beta}\delta_{1}g_{\mu \delta}\right)
-\frac{1}{2}g^{\gamma \beta}g^{\nu \mu}g^{\delta \alpha}\left(\delta_{2}g_{\nu \alpha}\nabla_{\beta}\delta_{1}g_{\mu \delta}\right)
-\frac{1}{2}g^{\gamma \nu}g^{\alpha \beta}g^{\mu \delta}\left(\delta_{2}g_{\nu \alpha}\nabla_{\beta}\delta_{1}g_{\mu \delta}\right)
\nonumber
\\
&\hskip 1 cm -\frac{1}{2}g^{\nu \alpha}g^{\gamma \mu}g^{\delta \beta}\left(\delta_{2}g_{\nu \alpha}\nabla_{\beta}\delta_{1}g_{\mu \delta}\right)
+\frac{1}{2}g^{\nu \alpha}g^{\gamma \beta}g^{\mu \delta}\left(\delta_{2}g_{\nu \alpha}\nabla_{\beta}\delta_{1}g_{\mu \delta}\right)
-\left(1\leftrightarrow 2\right) 
\nonumber
\\
&=g^{\gamma \mu}\left(g^{\delta \nu}g^{\alpha \beta}\delta_{2}g_{\nu \alpha}\right)\nabla_{\beta}\delta_{1}g_{\mu \delta}
-\frac{1}{2}\left(g^{\nu \mu}g^{\delta \alpha}\delta_{2}g_{\nu \alpha}\right)\nabla^{\gamma}\delta_{1}g_{\mu \delta}
-\frac{1}{2}\left(g^{\gamma \nu}g^{\alpha \beta}\delta_{2}g_{\nu \alpha}\right)\nabla_{\beta}\left(g^{\mu \delta}\delta_{1}g_{\mu \delta}\right)
\nonumber
\\
&\hskip 1 cm -\frac{1}{2}\left(g^{\nu \alpha}\delta_{2}g_{\nu \alpha}\right)\left(g^{\gamma \mu}g^{\delta \beta}\nabla_{\beta}\delta_{1}g_{\mu \delta}\right)
+\frac{1}{2}g^{\gamma \beta}\left(g^{\nu \alpha}\delta_{2}g_{\nu \alpha}\right)\left(g^{\mu \delta}\nabla_{\beta}\delta_{1}g_{\mu \delta}\right)
-\left(1\leftrightarrow 2\right)
\nonumber
\\
&=\frac{1}{2}\Bigg[\delta_{2}g^{\alpha \beta}\left(\nabla^{\gamma}\delta_{1}g_{\alpha \beta}\right)+\delta_{2}\ln g\left(\nabla_{\beta}\delta_{1}g^{\gamma \beta}\right)+\delta_{2}g^{\gamma \beta}\left(\nabla_{\beta}\delta_{1} \ln g\right)+\delta_{2}\ln g\left(\nabla^{\gamma}\delta_{1}\ln g \right)
\nonumber
\\
&\hskip 1 cm -2\delta_{2}g_{\alpha \beta}\left(\nabla^{\alpha}\delta_{1}g^{\gamma \beta}\right)-\left(1\leftrightarrow 2\right)\Bigg]
\label{omega_final1}
\end{align}
where we have simply raised and lowered the indices in a convenient form and have converted some terms to the determinant $g$ of the metric. In this form, each of the terms in $\omega^{\tau}$ can be easily evaluated. The following expressions are useful:
\begin{align}
\delta g_{\tau \tau}&= -\frac{2 \delta \sigma}{\tau}+o(1/\tau),  & \delta g^{\tau \tau} & = \frac{2 \delta \sigma}{\tau}+o(1/\tau),&
\\
\delta g_{ab}&=\tau \delta h^{(1)}_{ab}+o(\tau), &
\delta g^{ab}&=-\frac{1}{\tau^{3}}\delta h^{(1)ab}+o(1/\tau^{3}),&
\end{align}
and for the four-dimensional Christoffel symbols the following expressions are useful:
\begin{align}
& \Gamma^{c}_{\tau a} =\frac{1}{2}h^{cd}\partial_{\tau}h_{ad}=
\frac{1}{\tau}\delta^{c}_{a}+o(1/\tau), \\
&\Gamma^{\tau}_{\tau \tau}= \frac{1}{2} h^{\tau \tau} \partial_{\tau} h_{\tau \tau} =-\frac{\sigma}{\tau^{2}}+o(1/\tau^{2}),
\\
&\Gamma^{\tau}_{ab}=-\frac{1}{2} h^{\tau \tau} \partial_{\tau} h_{ab}=\tau h^{(0)}_{ab}+o(\tau).
\end{align}
Using these expressions, the first term in \ref{omega_final1} for $\gamma=\tau$ becomes,
\begin{align} \nonumber
\delta_{2}g^{\alpha \beta}\left(\nabla^{\tau}\delta_{1}g_{\alpha \beta}\right)&= \delta_{2} g^{\tau \tau}\left(\nabla^{\tau}\delta_{1}g_{\tau \tau}\right)+ \delta_{2} g^{ab}\left(\nabla^{\tau}\delta_{1}g_{ab}\right)
\\ 
\nonumber
&= -\bigg( \frac{2 \delta_{2} \sigma}{\tau}+\cdots\bigg)\bigg(\frac{2 \delta_{1}\sigma}{\tau^{2}}+\cdots\bigg)-\bigg( - \frac{\delta_{2} h^{(1)ab}}{\tau^{3}}+\cdots\bigg) \bigg(\delta_{1} h^{(1)}_{ab}+\cdots\bigg)
\\ 
\nonumber
&=\frac{1}{\tau^{3}}\left(\delta_{2}h^{(1)ab}\delta_{1}h^{(1)}_{ab}-4 \delta_{1} \sigma \delta_{2} \sigma\right)+o(1/\tau^{3})
\nonumber
\\
&=\frac{1}{\tau^{3}}\left(\delta_{2}k^{ab}\delta_{1}k_{ab}-2\delta_{2}\sigma \delta_{1}k-2\delta_{1}\sigma \delta_{2}k+8\delta_{1}\sigma \delta_{2}\sigma \right)+o(1/\tau^{3})
\end{align}
The second term becomes,
\begin{align} \nonumber
\delta_{2} \ln g \big(\nabla_{\beta} \delta_{1} g^{\tau \beta}\big)&=\bigg(\frac{2 \delta_{2} \sigma}{\tau}+\frac{h^{(0)}_{ab}\delta_{2}h^{(1)ab}}{\tau}+\cdots\bigg) \bigg(\frac{4 \delta_{1}\sigma}{\tau^{2}}-\frac{h^{(0)}_{ab}\delta_{1}h^{(1)ab}}{\tau^{2}}+\cdots\bigg) \\ 
\label{omega_4}
&=\frac{1}{\tau^{3}} \bigg(-40 \delta_{1}\sigma \delta_{2}\sigma+4 \delta_{1} k \delta_{2}\sigma+10 \delta_{2}k \delta_{1}\sigma- \delta_{1}k \delta_{2}k\bigg)+o(1/\tau^{3})
\end{align}
The third term becomes,
\begin{align} \nonumber
\delta_{2} g^{\tau \beta} \big(\nabla_{\beta} \delta_{1} \ln g\big) &=\delta_{2} g^{\tau \tau} ~\partial_{\tau} \big(\delta_{1} \ln g\big) \\ \nonumber
&=\bigg(\frac{2 \delta_{2}\sigma}{\tau} + \cdots\bigg)\bigg(-\frac{2 \delta_{1} \sigma}{\tau^{2}}- \frac{h^{(0)}_{ab}\delta_{1} h^{(1)ab}}{\tau^{2}}+\cdots\bigg)
\\ \label{omega_5}
&= \frac{1}{\tau^{3}} \big(8 \delta_{1}\sigma \delta_{2}\sigma - 2\delta_{2} \sigma \delta_{1} k \big) +o(1/\tau^{3})
\end{align}
The fourth term becomes
\begin{align} \nonumber 
  \delta_{2} \ln g \nabla^{\tau} \delta_{1} \ln g & = -\bigg( \frac{2 \delta_{2}\sigma}{\tau}+\frac{h^{(0)}_{ab}\delta_{2} h^{(1)ab}}{\tau}+\cdots\bigg)\partial_{\tau}\bigg( \frac{2 \delta_{1}\sigma}{\tau}+\frac{h^{(0)}_{ab}\delta_{1} h^{(1)ab}}{\tau}+\cdots\bigg) 
  \\ 
  &=\frac{1}{\tau^{3}} \bigg(16 \delta_{1} \sigma \delta_{2} \sigma - 4 \delta_{1} \sigma  \delta_{2} k- 4 \delta_{1}k \delta_{2} \sigma + \delta_{1} k \delta_{2} k\bigg) + o(1/\tau^{3})
\end{align}
The fifth term becomes
\begin{align}
\delta_{2}g_{\alpha \beta}\left(\nabla^{\alpha}\delta_{1}g^{\tau \beta}\right)&=\delta_{2}g_{\tau \tau}\left(\nabla^{\tau}\delta_{1}g^{\tau \tau}\right)+\delta_{2}g_{ab}\left(\nabla^{a}\delta_{1}g^{\tau b}\right)
\nonumber
\\
\nonumber
&=\left(\frac{2\delta_{2}\sigma}{\tau}+\cdots\right)\left(-\frac{2\delta_{1}\sigma}{\tau^{2}}+\cdots\right)+ \frac{2}{\tau^{3}}\bigg( \delta_{1} \sigma \delta_{2} h^{(1)}_{ab}h^{(0)ab}+\cdots\bigg)
\\  
&= \frac{1}{\tau^{3}} \bigg(2 \delta_{1} \sigma \delta_{2}k -16 \delta_{1} \sigma \delta_{2} \sigma\bigg)
+o(1/\tau^{3})
\end{align} 
Most of these terms cancel out upon $\left(1\leftrightarrow 2\right)$ anti-symmetrisation. The final expression for $\omega^{\tau}$ reads,
\begin{align} \label{omega_final}
\omega^{\tau}= \frac{2}{\tau^{3}}\big( \delta_{1} \sigma \delta_{2} k -  \delta_{1} k \delta_{2} \sigma\big)+o(1/\tau^{3})
    \end{align}
Using the boundary condition, $k=0$, the $\mathcal{O}(1/\tau^{3})$ term in \ref{omega_final} vanishes. 
Hence, in the $\tau \to \infty$ limit
\be
 \Omega(g,\delta_{1}g,\delta_{2}g) = 0.
\ee
This implies, 
\be
 \delta Q_\xi ( i^+) =0.
\ee
To summarise: we have shown that with our notion of asymptotic flatness, timelike infinity
does not contribute to the Lee-Wald symplectic form. Hence, the contribution to the first variations of the Iyer-Wald charges from timelike infinity is zero. It has been suggested by several authors that this should be the case. The result is entirely expected on physical grounds. \ref{Iyer-Wald-null-timelike} simplifies to 
\be
\delta Q_\xi (\Sigma) = \delta Q_\xi ( \mathcal{J}^+).
\ee

The contribution from null infinity, $\delta Q_\xi ( \mathcal{J}^+)$, is well studied; for a review see \cite{Ashtekar:2014zsa}. One of key ideas in the subject is that the integral over null infinity can be written as the difference of localised charges \cite{Chandrasekaran:2018aop}
\be
Q_\xi ( \mathcal{J}^+) = Q^{\rm{loc}}_\xi ( \mathcal{J}^+_-) - Q^{\rm{loc}}_\xi ( \mathcal{J}^+_+),
\ee
where $ \mathcal{J}^+_\pm$ are respectively the future and past 2-sphere limits of null infinity  $\mathcal{J}^+$. In the following we will be interested in $Q^{\rm{loc}}_\xi ( \mathcal{J}^+_+)$, which is what we call ``future charges''.  Timelike infinity  hyperboloid $\cal{H}$ reaches $\mathcal{J}^+_+$ in the $\rho \to \infty$ limit.

\subsection{Charges}
\label{sec:charges_timelike}

Since the contributions to the Lee-Wald symplectic form from timelike infinity vanishes, it follows that  “future charges” can be computed on any two-dimensional topologically-spherical surface surrounding the “sources” at timelike infinity. To keep the notation simple, we denote future charges by simply $Q_\xi $, instead of $Q^{\rm{loc}}_\xi ( \mathcal{J}^+_+)$.

Motivated by the corresponding expressions at spatial infinity \cite{Compere:2011ve}, we \textit{propose} expressions for  supertranslation and Lorentz charges at timelike infinity and show appropriate conservation properties. We do not present a first principal derivation for these expressions. Such a derivation can be given, for example, by relating the expressions below to the corresponding expressions to null infinity, but such a  calculation is not attempted in this work.

We begin by observing some elementary properties of the $1/\tau$ expansion of the Weyl tensor projected on $\tau=\textrm{constant}$ hypersurface. In four spacetime dimensions, the Weyl tensor expressed in terms of the Riemann tensor, Ricci tensor and Ricci scalar takes the form, 
\begin{align}\label{Weyl_Decomp_MT}
W_{\alpha \beta \mu \nu}=R_{\alpha \beta \mu \nu}-\frac{1}{2}\left(g_{\alpha \mu}R_{\beta \nu}+R_{\alpha \mu}g_{\beta \nu}-g_{\alpha \nu}R_{\beta \mu}-R_{\alpha \nu}g_{\beta \mu}\right)+\frac{R}{6}\left(g_{\alpha \mu}g_{\beta \nu}-g_{\alpha \nu}g_{\beta \mu} \right).
\end{align}
Let  $(\tau, \phi^a)$ be the four-dimensional spacetime coordinates associated to the 3+1 split. Then, for a general  spacetime coordinates $x^\mu= x^\mu(\tau, \phi^a)$ we define 
\be
e^{\mu}_{a} = \frac{\partial x^\mu}{\partial \phi^a}.
\ee
The vectors  $e^{\mu}_{a}$ with $\{a=1,2,3\}$ are tangent to  $\tau=\textrm{constant}$ hypersurface.  
The projected electric part of the Weyl tensor on $\tau=\textrm{constant}$ hypersurface is defined as,
\begin{align}
E_{ab}&=W_{\alpha \beta \mu \nu}e^{\alpha}_{a}n^{\beta}e^{\mu}_{b}n^{\nu}.
\end{align}
For vacuum spacetimes, with $R_{\alpha \beta}=0=R$, Gauss–Codazzi equations give,
\begin{align}
E_{ab}=R_{\alpha \beta \mu \nu}e^{\alpha}_{a}n^{\beta}e^{\mu}_{b}n^{\nu}=-\pounds_{n}K_{ab}+K_{ac}K^{c}_{b}+N^{-1}D_{a}D_{b}N~,
\end{align}
where $\pounds_{n}$ is the Lie-derivative with respect to the unit normal $n^\mu$.

Given the expansions for the extrinsic curvature components and the lapse function $N$ in powers of $1/\tau$, we can  obtain the expansion of the electric part of the Weyl tensor. A calculation gives,
\begin{align}
E_{ab} & \equiv \frac{1}{\tau}E_{ab}^{(1)}+\frac{1}{\tau^{2}}E_{ab}^{(2)} + \cdots,
\end{align}
where the zeroth order expansion coefficient identically vanishes and the first order expansion coefficient  is,
\begin{align}\label{First_electric_MT}
E_{ab}^{(1)}= \sigma_{ab} -\sigma h^{(0)}_{ab}.
\end{align}
The first order electric part of the Weyl tensor satisfies the following properties on $\mathcal{H}$,
\begin{align}
& E^{(1)}_{ab}=E^{(1)}_{ba}, &  & (\textrm{symmetric}) &
\\
& E^{(1)a}_{a}=\square \sigma -3\sigma =0, &  & (\textrm{traceless}) &
 \\
& \mathcal{D}_{b} E^{(1)b}_{a} = 0, & & (\textrm{divergence-free}) &
\end{align}
upon using the first order equations of motion.  It then follows that for conformal Killing vectors $\xi^a$ on $\mathcal{H}$, $E^{(1)}_{ab} \xi^a$ is a conserved current. The four translations induce four conformal Killing vectors $\xi^a = \mathcal{D}^a \omega$ on $\mathcal{H}$ (recall when $\omega$ represents a translation for $\omega_{ab} - h^{(0)}_{ab} \omega = 0$), and this conserved current can be used to construct ``future charges''~\cite{porrill, Gen:1997az},
\be
Q_\xi = -\frac{1}{8\pi G} \int_C \sqrt{q} \, d^2 x  \, E^{(1)}_{ab} \xi^a r^b
\ee 
where $C$ is a two dimensional topologically-spherical surface surrounding sources on $ \mathcal{H}$. The induced metric on $C$ is $q_{ab}$ and $r^a$ is the unit outward normal to $C$ in ${\cal H}$. These charges are ``conserved'' in the sense that the integral can be done on any topologically-spherical surface $C$ of $ \mathcal{H}$ surrounding the sources, and the answer is independent of the choice of $C$.

Clearly for supertranslations, such a construction \textit{does not} work as $\mathcal{D}^a \omega$ is not a conformal Killing vector on $ \mathcal{H}$.  Fortunately, a slight modification of this construction works \cite{Compere:2011ve}. We have,
\be
E^{(1)}_{ab} \xi^a = E^{(1)}_{ab} \omega^a =  \sigma_{ab}\omega^a - \sigma \omega_b.
\ee
Next consider $2 \mathcal{D}^a(\omega_{[a}\sigma_{b]} ) $ for translations $\omega_{ab} - h^{(0)}_{ab} \omega = 0$, i.e., 
\bea
2 \mathcal{D}^a(\omega_{[a}\sigma_{b]} ) &=& \mathcal{D}^a(\omega_{a} \sigma_b ) - \mathcal{D}^a(\omega_{b} \sigma_a) = 3 \omega \sigma_b + \omega_a \sigma_{b}{}^{a} - \omega_b{}^a \sigma_a - \omega_{b} (3\sigma) \\
&=& 3 \omega \sigma_b + \omega^a \sigma_{ab} - \omega \sigma_b - 3 \omega_{b} \sigma \\
&=&  \sigma_{ab} \omega^a  +  2 \omega \sigma_b  - 3 \sigma \omega_{b} 
\eea
Hence, for translations, 
\be
E^{(1)}_{ab} \xi^a - 2 \mathcal{D}^a(\omega_{[a}\sigma_{b]} ) = 2 (\sigma \omega_{b} - \omega \sigma_b).
\ee
The key point is that the term $\mathcal{D}^a(\omega_{[a}\sigma_{b]} )\xi^a r^b $ when integrated over $C$ only contributes a total divergence and therefore is zero. Hence, 
\be
 \int_C \sqrt{q} \, d^2 x  \, (E^{(1)}_{ab} \xi^a - 2 \mathcal{D}^a(\omega_{[a}\sigma_{b]} ) ) \xi^a r^b  =  \int_C \sqrt{q} \, d^2 x  \, E^{(1)}_{ab} \xi^a r^b =  2 \int_C \sqrt{q} \, d^2 x  \,  (\sigma \omega_{b} - \omega \sigma_b) r^b.
\ee
This last expression admits generalisation for supertranslations. The current $(\sigma \omega_{b} - \omega \sigma_b)$ is conserved for supertranslations as well, since $(\square - 3) \omega = 0$, implying 
\be
\mathcal{D}^b (\sigma \omega_{b} - \omega \sigma_b) = 0.
\ee
Hence, we can define a charge for supertranslation $\omega$ as
\be
Q_\omega = - \frac{1}{4\pi G} \int_C  \sqrt{q} \, d^2x \, (\sigma \omega_{b} - \omega \sigma_b) r^b.  \label{ST_charge}
\ee
For translations this expression reduces to the previous expressions~\cite{porrill, Gen:1997az}.

Expression for Lorentz charges is relativity easier to propose. One of the second order equation of motion, namely \ref{second_order_02}, automatically gives a conserved tensor,
\be
J_{ab} = -h^{(2)}_{ab} + \frac{1}{2} k_{a}^{c}k_{bc}+h^{(0)}_{ab}\left(-\frac{1}{8} k_{cd}k^{cd}+8 \sigma^{2}-k_{cd} \mathcal{D}^{c}\mathcal{D}^{d}\sigma -\mathcal{D}_{c}\sigma \mathcal{D}^{c}\sigma \right)
\ee
with $\mathcal{D}^a J_{ab} = 0$. For a Killing vector $\xi^a$ on $\mathcal{H}$ representing a four-dimensional rotation or boost we define, 
\be
Q_\xi = \frac{1}{8\pi G} \int_C \sqrt{q} d^2x \, J_{ab} \xi^a r^b.
\ee
These charges match with~\cite{porrill, Gen:1997az} upon setting $k_{ab} = 0$ and noting the fact that the second order magnetic part of the Weyl tensor is related to $J_{ab}$ by the curl operation defined in appendix \ref{Weyl_tensor_expansion}.

\subsection{Commutator of charges}
\label{sec:charges_commutator}
In the previous section, we wrote expressions for supertranslation and  Lorentz charges. The Poisson bracket between two charges is defined as (see e.g., \cite{Compere:2018aar, Chandrasekaran:2018aop}), 
\begin{align} \label{Poisson}
\{{Q}_{\chi},{Q}_{\chi'}\}=-\delta_{\chi}{Q}_{\chi'}
\end{align}
where the variation $\delta_{\chi}$ acts on the fields as the transformation induced by the asymptotic symmetry.
Supertranslation charges defined in \ref{ST_charge} can also be written as 
\begin{align} 
{Q}_{\omega}=\frac{1}{4\pi G}\int _{C}d^{2}x \, \sqrt{q} \, \left(\sigma_{a}\omega-\sigma \omega_{a}\right) r^{a}
=\frac{1}{4\pi G}\int _{\mathcal{V}}d^{3}x~\partial_{a}\left[\sqrt{-h^{(0)}}\left(\sigma^{a}\omega-\sigma \omega^{a}\right)\right],
\end{align}
where $\mathcal{V}$ is the part of  $\mathcal{H}$ surrounded by $C$. Now, we wish to compute the Poisson bracket between Lorentz charges and supertranslation charges.
Identifying $\chi=\xi$ (a Lorentz transformation) and $\chi'=\omega$ (a supertranslation), \ref{Poisson} becomes,
\begin{align} \label{Comm1}
\{{Q}_{\xi},{Q}_{\omega}\}&=-\delta_{\xi}{Q}_{\omega}~.
\end{align}
Using which the Poisson bracket becomes,
\begin{align}
\{{Q}_{\xi},{Q}_{\omega}\}&=-\frac{1}{4\pi G}\int _{\mathcal{V}}d^{3}x~\partial_{a}\left\{\delta_{\xi}\left[\sqrt{-h^{(0)}}\left(\sigma^{a}\omega-\sigma \omega^{a}\right)\right]\right\}
\nonumber
\\
&=-\frac{1}{4\pi G}\int _{\mathcal{V}}d^{3}x~\partial_{a}\left[\sqrt{-h^{(0)}}\left\{\xi^{b}\mathcal{D}_{b}\left(\sigma^{a}\omega-\sigma \omega^{a}\right)-\left(\sigma^{b}\omega-\sigma \omega^{b}\right)\mathcal{D}_{b}\xi^{a}\right\}\right]
\nonumber
\\
&=-\frac{1}{4\pi G}\int _{\mathcal{V}}d^{3}x~\partial_{a}\Big[\sqrt{-h^{(0)}}\Big\{\sigma^{a}\left(\xi^{b}\mathcal{D}_{b}\omega\right)+\omega \xi^{b}\mathcal{D}_{b}\sigma^{a}-\left(\xi^{b}\mathcal{D}_{b}\sigma\right)\omega^{a}-\sigma\left(\xi^{b}\mathcal{D}_{b}\omega^{a}\right)
\nonumber
\\
&\hskip 2 cm 
-\left(\sigma^{b}\omega-\sigma \omega^{b}\right)\mathcal{D}_{b}\xi^{a}\Big\}\Big]
\nonumber
\\
&=\frac{1}{4\pi G}\int _{\mathcal{V}}d^{3}x~\partial_{a}\Big[\sqrt{-h^{(0)}}\Big\{\sigma^{a}\left(-\xi^{b}\mathcal{D}_{b}\omega\right)-\sigma\left(-\xi^{b}\mathcal{D}_{b}\omega^{a}-\omega^{b}\mathcal{D}_{a}\xi_{b}\right) \Big\}\Big]
\nonumber
\\
&\hskip 2 cm -\frac{1}{4\pi G}\int _{\mathcal{V}}d^{3}x~\partial_{a}\Big[\sqrt{-h^{(0)}}\Big\{\omega \xi^{b}\mathcal{D}_{b}\sigma^{a}-\left(\xi^{b}\mathcal{D}_{b}\sigma\right)\omega^{a}-\sigma^{b}\omega \mathcal{ D}_{b}\xi^{a}\Big\}\Big]
\nonumber
\\
&=\frac{1}{4\pi G}\int _{\mathcal{V}}d^{3}x~\partial_{a}\Big[\sqrt{-h^{(0)}}\Big\{\sigma^{a}\left(\pounds_{-\xi}\omega\right)-\sigma \mathcal{D}^{a}\left(\pounds_{-\xi}\omega\right) \Big\}\Big]
\nonumber
\\
&\hskip 2 cm -\frac{1}{4\pi G}\int _{\mathcal{V}}d^{3}x~\partial_{a}\Big[\sqrt{-h^{(0)}}\Big\{\omega \xi^{b}\mathcal{D}_{b}\sigma^{a}-\left(\xi^{b}\mathcal{D}_{b}\sigma\right)\omega^{a}-\sigma^{b}\omega \mathcal{D}_{b}\xi^{a}\Big\}\Big]
\nonumber
\\
&={Q}_{\omega'}+\frac{1}{4\pi G}\int _{\mathcal{V}}d^{3}x~\sqrt{-h^{(0)}}\mathcal{D}_{a}\Big[\omega \xi^{b}\mathcal{D}_{b}\sigma^{a}-\left(\xi^{b}\mathcal{D}_{b}\sigma\right)\omega^{a}-\sigma^{b}\omega \mathcal{D}_{b}\xi^{a}\Big]
\nonumber
\\
&={Q}_{\omega'}+\frac{1}{4\pi G}\int _{\mathcal{V}}d^{3}x~\sqrt{-h^{(0)}}\Big[\omega \xi^{b}\mathcal{D}_{a}\mathcal{D}_{b}\sigma^{a}-\left(\xi^{b}\mathcal{D}_{b}\sigma\right)\square \omega-\sigma^{b}\omega \mathcal{D}_{a}\mathcal{D}_{b}\xi^{a}\Big]
\nonumber
\\
&={Q}_{\omega'}+\frac{1}{4\pi G}\int _{\mathcal{V}}d^{3}x~\sqrt{-h^{(0)}}\Big[\omega \xi^{b}[\mathcal{D}_{a},\mathcal{D}_{b}]\sigma^{a}+3\omega \xi^{b}\sigma_{b}-3\left(\xi^{b}\mathcal{D}_{b}\sigma\right)\omega-\sigma^{b}\omega [\mathcal{D}_{a},\mathcal{D}_{b}]\xi^{a}\Big]
\nonumber
\\
&={Q}_{\omega}+\frac{1}{4\pi G}\int _{\mathcal{V}}d^{3}x~\sqrt{-h^{(0)}}\Big[\omega \xi^{b}R^{(0)}_{ab}\sigma^{a}+3\omega \xi^{b}\sigma_{b}-3\left(\xi^{b}\mathcal{D}_{b}\sigma\right)\omega-\sigma^{b}\omega R_{ab}^{(0)}\xi^{a}\Big]
\nonumber
\\
&={Q}_{\omega'}
\end{align}
where $\omega'=\pounds_{-\xi}\omega$.
We have used the result, $\delta_{\xi}\sqrt{-h^{(0)}}=(1/2)\sqrt{-h^{(0)}}h^{(0)ab}\delta_{\xi}h^{(0)}_{ab}=0$.

One could attempt the calculation the other way round, i.e., identifying $\chi'=\xi$ (a Lorentz transformation) and $\chi=\omega$ (a supertranslation). That calculation is more involved. We expect to recover ${Q}_{\omega'}$ possibly with terms that only contribute to a total divergence on $C$. At spatial infinity the technology for 
identifying total divergence on the cuts of de Sitter hyperboloid is fairly well developed, see e.g.,~\cite{Mann:2008ay}; at timelike infinity some further technical work is required.

\section{The Schwarzschild solution near timelike infinity}
\label{sec:schw}
In this section, we write the Schwarzschild solution near timelike infinity in the Beig-Schmidt form \ref{Beig_Schmidt_form}--\ref{Beig_Schmidt_form2}. The Schwarzschild metric in standard static coordinates takes the form
\be
ds^2 = -\left(1 - \frac{2 G M}{r} \right) dt^2 + \left(1 - \frac{2 G M}{r} \right)^{-1} dr^2 + r^2  d\Omega^2 ,
\ee
where $d \Omega^2 = (d\theta^2 + \sin^2 \theta d \varphi^2)$ is the round metric on the unit two-sphere. We begin by introducing $(\tau_0, \rho_0)$ coordinates defined as follows:  
\bea
&& t = \tau_0 \sqrt{1 + \rho_0^2}, \\
&& r = \rho_0 \tau_0.
\eea
These coordinates do not bring the Schwarzschild solution near timelike infinity in the Beig-Schmidt form as in \ref{Beig_Schmidt_form}--\ref{Beig_Schmidt_form2}. A chain of further coordinate transformations outlined in appendix \ref{asymptotic_metric} are required (as expected). In coordinates $(\tau_0, \rho_0)$ the non-zero components of the metric takes the form to leading order in $1/\tau_0$:
\begin{align}
& g_{\tau_0 \tau_0} = - 1 + (2GM) \left(  \rho_0^{-1} + 2 \rho_0 \right) \frac{1}{\tau_0} + \mathcal{O}(\tau_0^{-2}) \\
& g_{\rho_0 \tau_0} = 4 GM +  \mathcal{O}(\tau_0^{-1}) \\
& g_{\rho_0 \rho_0} = \tau_0^2 (1 + \rho_0^2)^{-1} + (2GM) (1 + \rho_0^2)^{-1} \left(  \rho_0^{-1} + 2 \rho_0 \right) \tau_0 + \mathcal{O}(1) \\
& g_{\theta \theta} =   \rho_0^2  \tau_0^2  + \mathcal{O}(1) \\
& g_{\varphi \varphi} =   \rho_0^2  \tau_0^2 \sin^2 \theta + \mathcal{O}(1).
\end{align}
Since $g_{\rho_0 \tau_0}$ term does not fall-off as  $\mathcal{O}(\tau_0^{-1})$, the metric is not in the Beig-Schmidt form at $\mathcal{O}(\tau_0^{-1})$. To fix this, following appendix \ref{asymptotic_metric} we do the transformation,
\begin{align}
&\rho_0 = \rho_1 + \frac{G(\rho_1)}{\tau_1}, \\
&G(\rho_1) = 4 GM (1 + \rho_1^2), \\
&\tau_0 = \tau_1.
\end{align}
In the new coordinates $(\tau_1, \rho_1)$ the non-zero metric components take the form,
\begin{align}
& g_{\tau_1 \tau_1} = - 1 + (2GM) \left(  \rho_1^{-1} + 2 \rho_1 \right) \frac{1}{\tau_1} + \mathcal{O}(\tau_1^{-2}) \\
& g_{\rho_1 \tau_1} =   \mathcal{O}(\tau_1^{-1}) \\
& g_{\rho_1 \rho_1} = \tau_1^2 (1 + \rho_1^2)^{-1} + (2GM) (1 + \rho_1^2)^{-1} \left(  \rho_1^{-1} + 6 \rho_1 \right) \tau_1 + \mathcal{O}(1) \\
& g_{\theta \theta} =   \rho_1^2  \tau_1^2  +  8 GM  \rho_1 (1 + \rho_1^2) \tau_1 + \mathcal{O}(1)\\
& g_{\varphi \varphi} =   \rho_1^2  \tau_1^2 \sin^2 \theta  +  8 GM  \rho_1 (1 + \rho_1^2) \sin^2 \theta \tau_1 + \mathcal{O}(1).
\end{align}

The above metric is in the Beig-Schmidt form, though it does not satisfy our boundary condition $k=0$. To bring the metric in the requisite form, we do a general supertranslation and call the final coordinates $(\tau, \rho)$:
\begin{align}
& \tau_1 = \tau - F(\rho), \\
& \rho_1 = \rho + \frac{1+ \rho^2}{\tau} \partial_\rho F(\rho), \\
& F(\rho) = - GM \left(\rho + 2 \sqrt{1 + \rho^2} \sinh^{-1} \rho \right). 
\end{align}
 $F(\rho)$ \textit{does not} satisfy $\square F = 3 F$. 
The resulting metric is in the requisite Beig-Schmidt form at first order in the expansion in inverse powers of $\tau$, and
\be
h^{(1)}_{ab} = - 2 \sigma h^{(0)}_{ab}.
\ee
That is, not only $k=0$, but the full $k_{ab}$ is zero. The field $\sigma$ takes the value,
\be
\sigma = - (GM) \left(  \rho^{-1} + 2 \rho \right), \qquad \square \sigma = 3 \sigma.
\ee

From these transformations, we see that  as $\tau$ goes to $\infty$ for fixed $r$, $\rho$ goes to $0$. Thus, the horizon $r= 2 GM $  intersects the timelike infinity hyperboloid $\mathcal{H}$ at the origin $\rho =0$.  Note that the function $\sigma$ is singular at $\rho = 0$. 

The four functions satisfying 
\be
\mathcal{D}_a \mathcal{D}_b \omega - h^{(0)}_{ab} \omega = 0,
\ee
are $ \left\{ \sqrt{1 +\rho^2}, \rho \cos \theta, \rho \sin \theta \sin \phi,  \rho \sin \theta \cos \phi  \right \} $
representing respectively, the time-translation and three-spatial translations. The charge integral
\be
Q_\omega = - \frac{1}{4\pi G} \int_C  \sqrt{q} \, d^2x \, (\sigma \omega_{b} - \omega \sigma_b) r^b,
\ee
on $\rho = \textrm{constant}$ spherical surface $C$ for time-translation $\omega = \sqrt{1 +\rho^2}$ gives $M$.

\section{Some final remarks}
\label{sec:horizon}

In the previous section we saw that for the Schwarzschild solution the fields $\sigma$ and $ h^{(1)}_{ab} = -  2 \sigma h^{(0)}_{ab} $ are singular at $\rho=0$. The singularity is such that the charge integral is finite even on a $\rho=\epsilon$ surface $C$.  Thus, for the region $r > 2 GM$ of the Schwarzschild solution, timelike infinity is the hyperboloid $\mathcal{H}$ minus the origin. This indicates that for a system composed of individually bound systems, gravitationally  unbound relative to each other, timelike infinity for the spacetime region describing outside the world-tubes of these system can be taken to be  $\mathcal{H}$ minus one point each for the individually bound system. These points act as sources for the charge integrals.

\begin{figure}[t]
\begin{center}
 \includegraphics[width=0.6\textwidth]{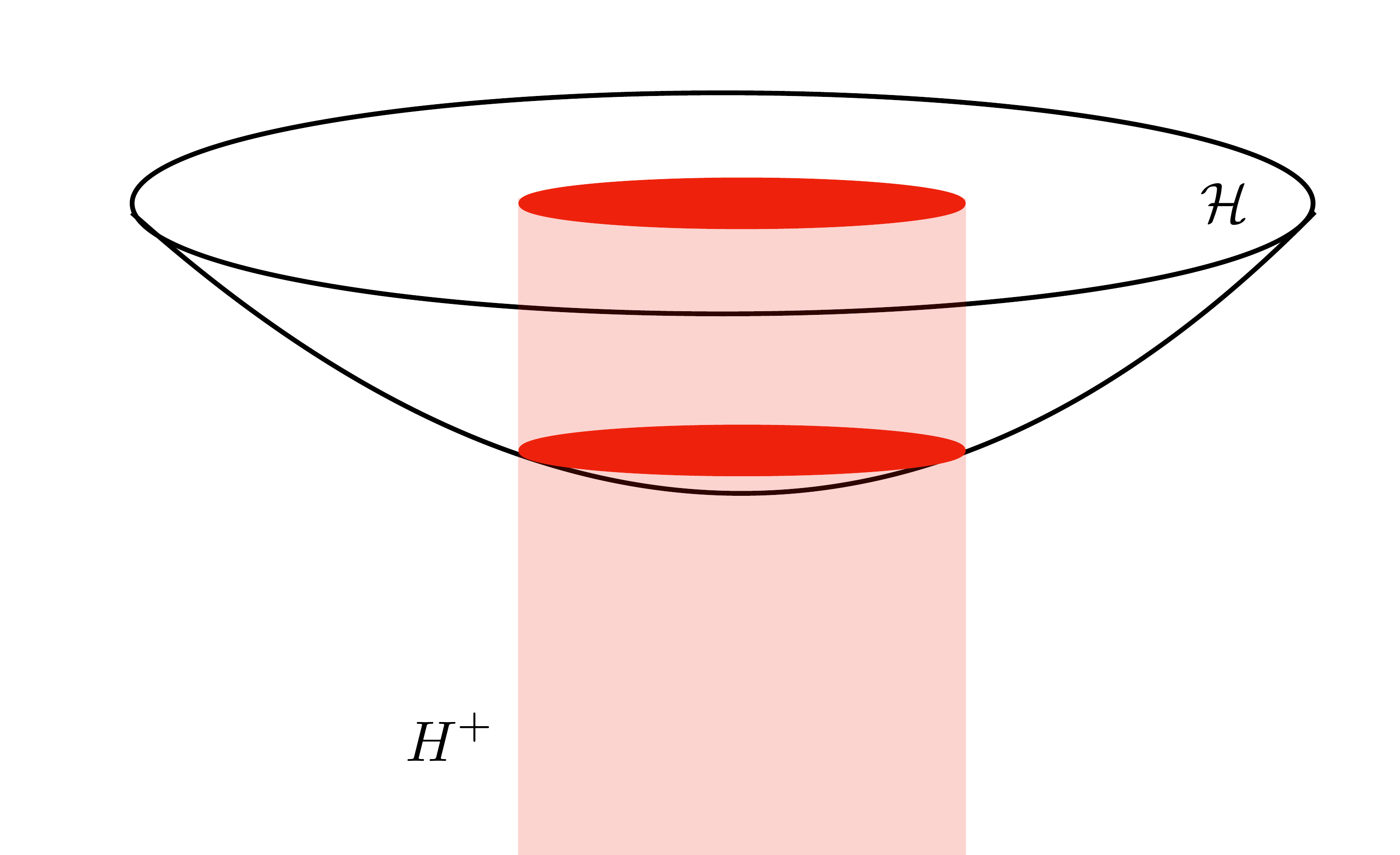}
 \caption{\sf Horizon ${H}^+$ intersecting the timelike infinity hyperboloid $\mathcal{H}$. In the limit $\tau \to \infty$ the intersection shrinks to a point. }
\label{fig:figure_timelike_infinity}
\end{center}
\end{figure}

\begin{figure}[t]
\begin{center}
 \includegraphics[width=0.4\textwidth]{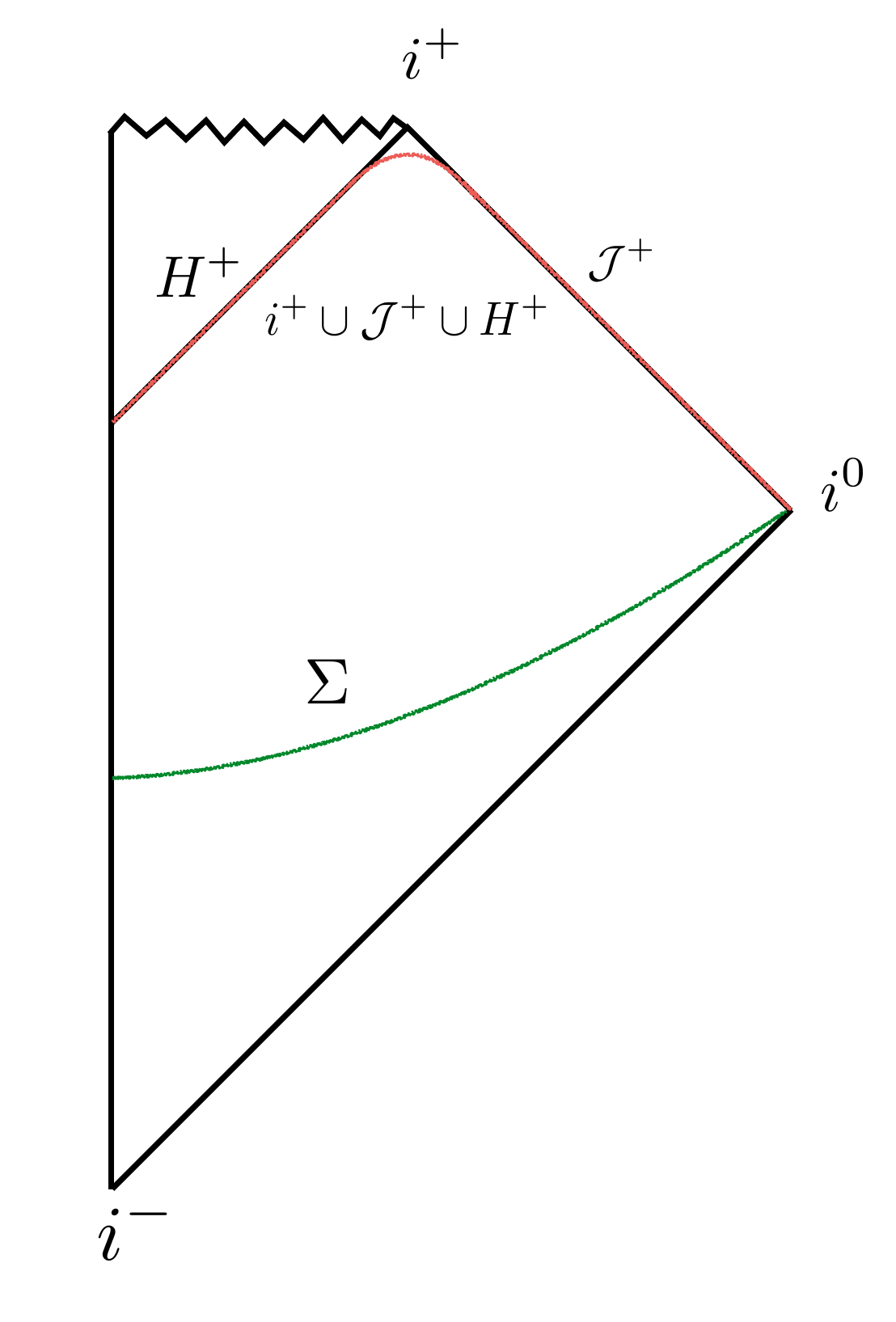}
 \caption{\sf For a black hole formed by gravitational collapse, components of the boundary are $\mathcal{J}^{-}$, $\mathcal{J}^{+}$, $H^{+}$ and the points at infinity $i^{-}$, $i^{0}$, and $i^{+}$. Since the first variation of global charges  is invariant under local deformations of the Cauchy surface $\Sigma$, one can deform  $\Sigma$ to $\mathcal{J}^+ \, \cup \,  i^+ \, \cup \, H^+$  in the far future. Then, $\delta Q_\xi (\Sigma) = \delta Q_\xi ( \mathcal{J}^+) + \delta Q_\xi ( i^+) +  \delta Q_\xi ( H^+)$. With our boundary conditions $\delta Q_\xi ( i^+) =0$. }
\label{fig:figure_flat_horizon}
\end{center}
\end{figure}

For simplicity we focus on only one bound system, represented as a black hole, and take the horizon to intersect the timelike infinity hyperboloid $\mathcal{H}$ at the origin $\rho =0$. We excise the point $\rho =0$: $i^+ =  \mathcal{H} \backslash  \{\rho =0 \}$. The horizon is a blow up of the point  $\rho = 0$ as schematically shown in~\ref{fig:figure_timelike_infinity}. Having excised the point $\rho=0$, the fields are all smooth at timelike infinity. The considerations of section \ref{sec:charges} can be carried over. The first variation of the Iyer-Wald charges at timelike infinity vanishes
\be
\delta Q_\xi (i^+) = 0.
\ee
This is schematically shown in figure \ref{fig:figure_flat_horizon}.

Let us comment on the general form of the solutions for $\omega$, $\sigma$ and relate it to the Green's function discussion of \cite{Campiglia:2015lxa}.
The supertranslation function $\omega$ and the field $\sigma$ both satisfy the equation $(\square - 3) f = 0$. Expanding in spherical harmonics, we have
\be
f(\rho, \theta, \varphi) = \sum_{l = 0}^{\infty}  \sum_{m=-l}^{l} f_l(\rho) Y_{lm} (\theta, \varphi).
\ee
The equation for functions $f_l(\rho)$ admits two classes of solutions. 
The first set takes the form, 
\be
f^{(I)}_l (\rho) = \frac{\rho ^l}{c_l}  \, _2F_1\left(\frac{l-1}{2},\frac{3 + l}{2};\frac{3}{2} + l ;-\rho ^2\right),
\ee
where $_2F_1$ is the standard hypergeometric function where $c_l = \frac{\Gamma\left(l + \frac{3}{2}\right)}{\Gamma\left(2+ \frac{l}{2}\right) \Gamma\left(\frac{3+ l}{2}\right)}$ is a convenient normalisation. In the $\rho \to 0$ limit these solutions go as 
$f^{(I)}_l (\rho) \sim \frac{1}{c_l} \rho^l$. In the  $\rho \to \infty$ limit they behave as 
$f^{(I)}_l (\rho) \sim \rho$. These functions correspond to supertranslations:
\be
 \omega (\rho, \theta, \varphi) =\sum_{l = 0}^{\infty}  \sum_{m=-l}^{l}  c_{lm}  f^{(I)}_l (\rho)  Y_{lm} (\theta, \varphi).
 \label{omega_ST}
 \ee
 This can be seen as follows. For Minkowski space, in outgoing coordinates $(u, r, \theta, \varphi) = (t - r, r, \theta, \varphi)$, the time-translation takes the form,
\be
\partial_u = \partial_t = \frac{\partial \tau }{\partial t}\partial_\tau + \frac{\partial \rho }{\partial t} \partial_\rho  = \sqrt{1 + \rho^2} \,  \partial_\tau - \frac{\rho \sqrt{1 + \rho^2}}{\tau} \, \partial_\rho.
\ee
In the  $\tau \to \infty$ limit and then $\rho \to \infty$ limit, $ \partial_u \sim  \rho \,  \partial_\tau$. Thus the expected  behaviour of $f(\theta, \varphi)  \partial_u$ is indeed the one captured by the supertranslations \ref{first_order_super}--\ref{first_order_super2} with $\omega (\rho, \theta, \varphi)$ given in \ref{omega_ST}. A general null infinity supertranslation  $f(\theta, \varphi)  \partial_u$ correspond to 
\be
f(\theta, \varphi) \partial_u = \sum_{l = 0}^{\infty}  \sum_{m=-l}^{l} c_{lm}  Y_{lm} (\theta, \varphi) \partial_u.
\ee
This construction, from the function $f(\theta, \varphi)$ to   $\omega (\rho, \theta, \varphi)$ via \ref{omega_ST}, is the same as the Green's function construction of reference \cite{Campiglia:2015lxa}.\footnote{A proof can be explicitly written using the addition theorem of spherical harmonics.}

The second independent set of solutions for the functions $f_l(\rho)$ takes the form
\be
f^{(II)}_l (\rho) = \rho ^{-l-1} \,
   _2F_1\left(-1 +\frac{l}{2},1 -\frac{l}{2};\frac{1}{2}-l;-\rho
   ^2\right).
\ee
In the $\rho \to 0$ limit these solutions go as 
$f^{(II)}_l (\rho) \sim \rho ^{-l-1}$. Explicitly first few of these functions are
\bea
f^{(II)}_0 (\rho) &=& \rho^{-1} + 2 \rho \\
f^{(II)}_1 (\rho) &=&\rho^{-2} (1-2 \rho^2) \sqrt{1+ \rho^2}, \\
f^{(II)}_2 (\rho) &=& \rho^{-3},
\eea
etc. For $l > 2$, in the  $\rho \to \infty$ limit they behave as 
$f^{(II)}_l (\rho) \sim~\textrm{const}~\rho^{-3}$. Our $\sigma$ for the Schwarzschild solution matches with $f^{(II)}_0 (\rho)$. Motivated by the corresponding discussion at spatial infinity, it is natural to speculate that the most general $\sigma$ consists of the linear sum of the functions $f^{(II)}_l (\rho) Y_{lm} (\theta, \varphi)$
\be
\sigma (\rho, \theta, \varphi) =\sum_{l = 0}^{\infty}  \sum_{m=-l}^{l}  d_{lm}  f^{(II)}_l (\rho)  Y_{lm} (\theta, \varphi).
\ee
Note that such a $\sigma$ is singular at $\rho =0$.

\section{Conclusions}
\label{sec:disc}
In this paper, we have initiated the study of supertranslations at timelike infinity. Largely developing on the previous works at spatial infinity, we have proposed a definition of asymptotic flatness at timelike infinity in four spacetime dimensions. We presented a thorough study of the asymptotic equations of motion and the action of supertranslations on asymptotic fields. We showed that the Lee-Wald symplectic form $\Omega (g, \delta_1 g, \delta_2 g)$ does not get contributions from the future timelike infinity with our boundary conditions.  As a result, the ``future charges'' can be computed on any two-dimensional surface surrounding the sources at timelike infinity. We presented expressions for supertranslation and Lorentz charges.  For general spacetimes we expect
\be
\textrm{future charges}~\xleftarrow{u \to + \infty}~\textrm{Bondi charges at}~\mathcal{J}^+~\xrightarrow{u \to - \infty}~\textrm{spatial infinity charges}.
\ee
Whether radiative spacetimes with non-trivial supertranslation charges exist that satisfy this hierarchy is open to argument \cite{Ashtekar:2018lor}.

Our work offers several opportunities for future research. We list a few directions. 

It is very much desirable to understand the relation between timelike infinity and  null infinity.  We expect our charge expressions can be matched with appropriate expressions  for supertranslation and Lorentz charges at $\mathcal{J}^+_+$ (the future endpoint of the future null infinity) following \cite{Troessaert:2017jcm, Prabhu:2019fsp}.

Can our boundary conditions we used to give a prescription for relating supertranslations at future null infinity to supertranslations at the horizon, thereby making the general idea mentioned in section 7 of \cite{Chandrasekaran:2018aop} more precise? Note that this viewpoint differs from that of \cite{Hawking:2016sgy} where global Bondi coordinates were used to link generators at the past null infinity $\mathcal{J}^-$ and the future horizon $H^+$.

Finally, there are other classes of transformations, e.g., logarithmic translations, superrotations, more general spi-supertranslations etc.~that we have not 
considered in this work. One would like to understand their action/role at timelike infinity. For much of our non-linear analysis we used the boundary condition $k =0$. Is it desirable to relax this condition?  

We hope to return to some of these problems in our future work.

\subsection*{Acknowledgements}
It is a pleasure to thank Abhay Ashtekar,  Miguel Campiglia, Alok Laddha, and C\'edric Troessaert for discussions on these topics over the years. The work of AV, AK, and DG  was supported in part by the Max Planck Partnergroup ``Quantum Black Holes'' between CMI Chennai and AEI Potsdam and by a grant to CMI from the Infosys Foundation. The work of SJH is supported in part by the Czech Science Foundation Grant 19-01850S. Research of SC is funded by the INSPIRE Faculty fellowship from DST, Government of India (Reg.~No.~DST/INSPIRE/04/2018/000893) and by the Start-Up Research Grant from SERB, DST, Government of India (Reg.~No.~SRG/2020/000409).

\appendix

\section{Asymptotic form of the metric}
 \label{asymptotic_metric}

We begin by considering a general class of  spacetimes admitting an expansion at timelike infinity of the form
\begin{align}\label{asymp_metric_01_APP}
g_{\mu \nu}=\eta_{\mu \nu}+\sum_{n=1}^{m}\ell_{\mu \nu}^{(n)}\left(\frac{x^{\sigma}}{\tau}\right)\frac{1}{\tau^{n}}+\cdots~,
\quad
\end{align}
where
\be
\tau^{2}=-\eta_{\mu \nu}x^{\mu}x^{\nu}, \label{asymp_metric_def_tau}
\ee
and where $x^{\mu}$ are a set of Cartesian coordinates on flat spacetime at infinity. This class of the spacetimes can be put in a more convenient form as in \ref{Beig_Schmidt_form}--\ref{Beig_Schmidt_form2}. In this appendix we do so explicitly, following Beig and Schmidt \cite{BeigSchmidt}. The form \ref{Beig_Schmidt_form}--\ref{Beig_Schmidt_form2} is our starting point for defining asymptotically flat spacetimes at timelike infinity.

The ten functions in $\ell_{\mu \nu}^{(n)}$ at any given order $n$ are functions of the dimensionless coordinate $(x^{\sigma}/\tau)$. To avoid cumbersome notation, henceforth in all the expressions we shall simply write $\ell_{\mu \nu}^{(n)}$ without mentioning its dependence on $(x^{\sigma}/\tau)$.

Instead of the Cartesian coordinates $x^{\mu}$, it is more convenient  to use $(\tau,\phi^{a})$ as a new set of coordinates, with $\tau$ defined in \ref{asymp_metric_def_tau} and $\phi^a$ are coordinates on hyperboloid $\cal{H}$. For any set of $\phi^a$ we define functions $\omega^{\mu}(\phi^{a})$, such that, 
\be
\omega^{\mu}(\phi^{a}) = \frac{x^{\mu}}{\tau}. 
\ee
Using this relation we get,
\be
dx^{\mu}=\omega^{\mu}d\tau+\tau(\partial_{a}\omega^{\mu})d\phi^{a}.
\ee 
Inserting the above equation in \ref{asymp_metric_01_APP} we obtain the following expression for the line element,
\begin{align}
ds^{2}&=g_{\mu \nu}dx^{\mu}dx^{\nu}=\left[\eta_{\mu \nu}+\sum_{n=1}^{m}\ell_{\mu \nu}^{(n)}\frac{1}{\tau^{n}}+\cdots\right]dx^{\mu}dx^{\nu}
\nonumber
\\
&=\left[\eta_{\mu \nu}+\sum_{n=1}^{m}\ell_{\mu \nu}^{(n)}\frac{1}{\tau^{n}}+\cdots\right]\left(\omega^{\mu}d\tau+\tau(\partial_{a}\omega^{\mu})d\phi^{a}\right)\left(\omega^{\nu}d\tau+\tau(\partial_{b}\omega^{\nu})d\phi^{b}\right)
\nonumber
\\
&=\left[\eta_{\mu \nu}+\sum_{n=1}^{m}\ell_{\mu \nu}^{(n)}\frac{1}{\tau^{n}}+\cdots\right]
\nonumber
\\
&\hskip 1 cm \times \left[\omega^{\mu}\omega^{\nu}d\tau^{2}+\tau\omega^{\nu}(\partial_{a}\omega^{\mu})d\phi^{a}d\tau+\tau\omega^{\mu}(\partial_{a}\omega^{\nu})d\phi^{a}d\tau+\tau^{2}(\partial_{a}\omega^{\mu})(\partial_{b}\omega^{\nu})d\phi^{a}d\phi^{b} \right]
\nonumber
\\
&=-\left[-\eta_{\mu \nu}\omega^{\mu}\omega^{\nu}-\sum_{n=1}^{m}\ell_{\mu \nu}^{(n)}\omega^{\mu}\omega^{\nu}\frac{1}{\tau^{n}}+\cdots \right]d\tau^{2}
\nonumber
\\
& \quad +2\tau \left[\eta_{\mu \nu}\omega^{\mu}(\partial_{a}\omega^{\nu})+\sum_{n=1}^{m}\ell_{\mu \nu}^{(n)}\omega^{\mu}(\partial_{a}\omega^{\nu})\frac{1}{\tau^{n}}  + \cdots \right] d\tau d\phi^{a}
\nonumber
\\
& \quad +\tau^{2}\left[\eta_{\mu \nu}(\partial_{a}\omega^{\mu})(\partial_{b}\omega^{\nu})+\sum_{n=1}^{m}\ell_{\mu \nu}^{(n)}(\partial_{a}\omega^{\mu})(\partial_{b}\omega^{\nu})\frac{1}{\tau^{n}} + \cdots \right]d\phi^{a}d\phi^{b}.
\end{align}
Using $\eta_{\mu \nu}\omega^{\mu}\omega^{\nu}=-1$ and $\eta_{\mu \nu}\omega^{\mu}(\partial_{a}\omega^{\nu})=(1/2)\partial_{a}(\omega_{\mu}\omega^{\mu})=0$, along with the following definitions,
\begin{align}
&\bar{\sigma}^{(n)}(\phi^{c})\equiv -\ell_{\mu \nu}^{(n)}\omega^{\mu}\omega^{\nu}~,\\
&A_{a}^{(n)}(\phi^{c}) \equiv \ell_{\mu \nu}^{(n)}\omega^{\mu}(\partial_{a}\omega^{\nu})~,\\
&h_{ab}^{(n)}(\phi^{c}) \equiv \ell_{\mu \nu}^{(n)}(\partial_{a}\omega^{\mu})(\partial_{b}\omega^{\nu})~\\ 
&h_{ab}^{(0)}(\phi^{c}) \equiv \eta_{\mu \nu}(\partial_{a}\omega^{\mu})(\partial_{b}\omega^{\nu})~,
\end{align}
the asymptotic form of the line element at timelike infinity takes the form,
\begin{align}
ds^{2}&=-\left[1+\sum_{n=1}^{m}\frac{\bar{\sigma}^{(n)}(\phi^{c})}{\tau^{n}}+\mathcal{O}\left(\tau^{-m-1}\right)\right]d\tau^{2}+2\tau \left[\sum_{n=1}^{m}\frac{A_{a}^{(n)}(\phi^{c})}{\tau^{n}}+\mathcal{O}\left(\tau^{-m-1}\right) \right] d\tau d\phi^{a}
\nonumber
\\
&\quad +\tau^{2}\left[h^{(0)}_{ab}+\sum_{n=1}^{m}\frac{h_{ab}^{(n)}(\phi^{c})}{\tau^{n}}+\mathcal{O}\left(\tau^{-m-1}\right) \right]d\phi^{a}d\phi^{b}~,
\\
&=-\left[1+\sum_{n=1}^{m}\frac{\sigma^{(n)}(\phi^{c})}{\tau^{n}}+\mathcal{O}\left(\tau^{-m-1}\right)\right]^{2}d\tau^{2}+2\tau \left[\sum_{n=1}^{m}\frac{A_{a}^{(n)}(\phi^{c})}{\tau^{n}}+\mathcal{O}\left(\tau^{-m-1}\right) \right] d\tau d\phi^{a}
\nonumber
\\
&\quad +\tau^{2}\left[h^{(0)}_{ab}+\sum_{n=1}^{m}\frac{h_{ab}^{(n)}(\phi^{c})}{\tau^{n}}+\mathcal{O}\left(\tau^{-m-1}\right) \right]d\phi^{a}d\phi^{b}~.
\label{asymp_met_final}
\end{align}
Here, $\sigma^{(n)}(\phi^{c})$ are functions of $\bar{\sigma}^{(n)}(\phi^{c})$, e.g., $\sigma^{(1)}=(\bar{\sigma}^{(1)}/2)$.

Next we show that there exist a coordinate transformation that brings the metric in \ref{asymp_met_final} to a form where, 
\begin{align}\label{coord_tran1}
&\sigma^{(n)}(\phi^{c})=0~, \quad \textrm{for}~ n\geq 2, \\
& A^{(n)}_{a}(\phi^{c})=0, \quad \textrm{for}~ n\geq 1. 
\label{coord_tran2}
\end{align}

We achieve this order by order. At first order, we take
\begin{align}
&\phi^{a}=\bar{\phi}^{a}+\frac{G^{(1)a}(\bar{\phi}^{b})}{\bar{\tau}},\\
&\tau=\bar{\tau}~.
\end{align}
This yields,
\be
d\phi^{a} =d\bar{\phi}^{a}-\frac{G^{(1)a}}{\bar{\tau}^{2}}d\bar{\tau}+\frac{1}{\bar{\tau}}\left(\partial_{b}G^{(1)a}\right)d\bar{\phi}^{b},
\ee
and
\be
\sigma^{(n)}(\phi^{c})=\sigma^{(n)}(\bar{\phi}^{c})+\frac{G^{(1)a}(\bar{\phi}^{b})}{\bar{\tau}}\partial_{a}\sigma^{(n)}+\mathcal{O}(\bar{\tau}^{-2}).
\ee
In these new coordinates, line element \ref{asymp_met_final} takes the form (keeping track of all the first order terms),
\begin{align}
ds^{2}&
 =-\left[1+\frac{2\sigma^{(1)}(\bar{\phi}^{c})}{\bar{\tau}}+\mathcal{O}\left(\bar{\tau}^{-2}\right)\right]d\bar{\tau}^{2}+2\bar{\tau} \left[\frac{A_{a}^{(1)}(\bar{\phi}^{c})}{\bar{\tau}}+\mathcal{O}\left(\bar{\tau}^{-2}\right) \right] d\bar{\tau}d\bar{\phi}^{a}
\nonumber
\\
&\quad +\bar{\tau}^{2}\left[h^{(0)}_{ab}+\frac{h_{ab}^{(1)}(\bar{\phi}^{c})}{\bar{\tau}}+\mathcal{O}\left(\tau^{-2}\right) \right]d\bar{\phi}^{a}d\bar{\phi}^{b}-2\bar{\tau}^{2}\left[h^{(0)}_{ab}+\frac{h_{ab}^{(1)}(\bar{\phi}^{c})}{\bar{\tau}}+\mathcal{O}\left(\tau^{-2}\right) \right]\frac{G^{(1)a}}{\bar{\tau}^{2}}d\bar{\tau}d\bar{\phi}^{b}
\nonumber
\\
&\quad +2\bar{\tau}^{2}\left[h^{(0)}_{ab}+\frac{h_{ab}^{(1)}(\bar{\phi}^{c})}{\bar{\tau}}+\mathcal{O}\left(\tau^{-2}\right) \right]\frac{1}{\bar{\tau}}\left(\partial_{c}G^{(1)a}\right)d\bar{\phi}^{c}d\bar{\phi}^{b}
\nonumber
\\
&=-\left[1+\frac{2\sigma^{(1)}(\bar{\phi}^{c})}{\bar{\tau}}+\mathcal{O}\left(\bar{\tau}^{-2}\right)\right]d\bar{\tau}^{2}
+2\left[A_{a}^{(1)}(\bar{\phi}^{c})-h^{(0)}_{ab}G^{(1)b}+\mathcal{O}\left(\bar{\tau}^{-1}\right) \right] d\bar{\tau}d\bar{\phi}^{a}
\nonumber
\\
& \quad +\bar{\tau}^{2}\left[h^{(0)}_{ab}+\frac{h_{ab}^{(1)}(\bar{\phi}^{c})}{\bar{\tau}}+\frac{2}{\bar{\tau}}h^{(0)}_{cb}\left(\partial_{a}G^{(1)c}\right)+\mathcal{O}\left(\tau^{-2}\right) \right]d\bar{\phi}^{a}d\bar{\phi}^{b}~.
\end{align}
Thus setting,
\begin{align}
A_{a}^{(1)}=h^{(0)}_{ab}G^{(1)b}~,
\end{align}
the line element takes the requisite form at the first order in the inverse powers of $\tau$.

Keeping track of the second order terms, we have
\begin{align}
ds^{2}&=-\left[1+\frac{2\sigma^{(1)}(\phi^{c})}{\tau}+\frac{(\sigma^{1})^{2}+2\sigma^{(2)}}{\tau^{2}}+\mathcal{O}\left(\tau^{-3}\right)\right]d\tau^{2}+2\tau \left[\frac{A_{a}^{(2)}(\phi^{c})}{\tau^{2}}+\mathcal{O}\left(\tau^{-3}\right) \right] d\tau d\phi^{a}
\nonumber
\\
&\quad  +\tau^{2}\left[h^{(0)}_{ab}+\frac{h_{ab}^{(1)}(\phi^{c})}{\tau}+\frac{h_{ab}^{(2)}(\phi^{c})}{\tau^{2}}+\mathcal{O}\left(\tau^{-3}\right) \right]d\phi^{a}d\phi^{b}~.
\label{asymp_met_final2}
\end{align}
The following coordinate transformation, 
\begin{align}\label{coordinate_transform}
& \tau=\bar{\tau}+\frac{F^{(2)}(\phi^{c})}{\bar{\tau}}, \\
&\bar{\phi}^{a}=\phi^{a},
\end{align}
yields
 \begin{align}
ds^{2}
 &=-\left[1+\frac{2\sigma^{(1)}}{\bar{\tau}}+\frac{(\sigma^{1})^{2}+2\sigma^{(2)}}{\bar{\tau}^{2}}-\frac{2F^{(2)}}{\bar{\tau}^{2}}+\mathcal{O}\left(\bar{\tau}^{-3}\right)\right]d\bar{\tau}^{2}+2\bar{\tau} \left[\frac{A_{a}^{(2)}-\partial_{a}F^{(2)}}{\bar{\tau}^{2}}+\mathcal{O}\left(\bar{\tau}^{-3}\right) \right] d\bar{\tau}d\phi^{a}
\nonumber
\\
&\quad +\bar{\tau}^{2}\left[h^{(0)}_{ab}+\frac{h_{ab}^{(1)}}{\bar{\tau}}+\frac{h_{ab}^{(2)}}{\bar{\tau}^{2}}+\frac{2F^{(2)}}{\bar{\tau}^{2}}h^{(0)}_{ab}+\mathcal{O}\left(\bar{\tau}^{-3}\right) \right]d\phi^{a}d\phi^{b}~.
\end{align}
Thus, no $(1/\bar{\tau})$ term has been generated in the coefficient of the $d\bar{\tau}d\phi^{a}$ term and hence the condition $A_{a}^{(1)}=0$ continues to hold. Furthermore, if we choose, 
\begin{align}
2F^{(2)}=2\sigma^{(2)}~,
\end{align}
then the $(1/\bar{\tau}^{2})$ term in the coefficient of the $d\bar{\tau}^{2}$ in the metric can be set to $(\sigma^{1})^{2}$. Thus the modified metric has only $[(\sigma^{1})^{2}/\bar{\tau}^{2}]$ term in the coefficient of the $d\bar{\tau}^{2}$ and no $(1/\bar{\tau})$ term in the coefficient of the $d\bar{\tau}d\phi^{a}$. We can now use, 
\begin{align}
&\phi^{a}=\bar{\phi}^{a}+\frac{G^{(2)a}(\bar{\phi}^{b})}{\bar{\tau}^{2}},\\
&\tau=\bar{\tau},
\end{align}
and choose the function $G^{(2)a}(\bar{\phi}^{b})$, such that $A^{(2)}_{a}h^{(0)ab}=G^{(2)b}$ and hence the $(1/\bar{\tau}^{2})$ term in the coefficient of the $d\bar{\tau}d\phi^{a}$ can be made to vanishes. Next, setting
\begin{align}
\tau=\bar{\tau}+\frac{F^{(3)}(\phi^{c})}{\bar{\tau}^{2}}
\end{align}
we can eliminate $(1/\bar{\tau}^{3})$ term in the coefficient of the $d\bar{\tau}^{2}$.

Proceeding in an identical manner, we can eliminate all terms in the coefficient of  $d\bar{\tau}d\phi^{a}$ and all terms beyond $\bar{\tau}^{-2}$ in the coefficient of $d\bar{\tau}^{2}$. Thus, metric \ref{asymp_met_final} can be reduced to the one satisfying conditions \ref{coord_tran1}--\ref{coord_tran2}. Thereby, we arrive at our final form \ref{Beig_Schmidt_form}--\ref{Beig_Schmidt_form2}.

\section{Action of supertranslation on asymptotic fields}
\label{supertranslations_app}
We apply the following transformation,
\begin{align}
&\quad \tau = \bar{\tau}-\omega(\bar{\phi}^{a})+\frac{1}{\bar{\tau}}F^{(2)}(\bar{\phi}^{a})+\mathcal{O}\left(\frac{1}{\bar{\tau}^{2}}\right)~,
\\
&\quad \phi^{a} = \bar{\phi}^{a}+\frac{1}{\bar{\tau}}h^{(0)ab}\partial_{b}\omega(\bar{\phi}^{c})+\frac{1}{\bar{\tau}^{2}}G^{(2)~a}(\bar{\phi}^{c})+\mathcal{O}\left(\frac{1}{\bar{\tau}^{3}}\right)~.
\end{align}
We obtain, 
\begin{align}
d\tau = & \left[1-\frac{1}{\bar{\tau}^{2}}F^{(2)}+\mathcal{O}\left(\frac{1}{\bar{\tau}^{3}}\right)\right]d\bar{\tau}+\left[-\partial_{a}\omega+\frac{1}{\bar{\tau}}\partial_{a}F^{(2)}+\mathcal{O}\left(\frac{1}{\bar{\tau}^{2}}\right)\right]d\bar{\phi}^{a}~,
\\
d\phi^{a}= & \left[\delta^{a}_{c}+\frac{1}{\bar{\tau}}\partial_{c}\left(h^{(0)ab}\partial_{b}\omega\right)+\frac{1}{\bar{\tau}^{2}}\left(\partial_{c}G^{(2)a}\right)+\mathcal{O}\left(\frac{1}{\bar{\tau}^{3}}\right)\right]d\bar{\phi}^{c}
\nonumber
\\
&   +\left[-\frac{1}{\bar{\tau}^{2}}\left(h^{(0)ab}\partial_{b}\omega\right)-\frac{2}{\bar{\tau}^{3}}G^{(2)a}+\mathcal{O}\left(\frac{1}{\bar{\tau}^{4}}\right)\right]d\bar{\tau}~.
\end{align}
The following relations are also obtained,
\begin{align}
 & \sigma(\phi^{a}) =\sigma(\bar{\phi}^{a})+\frac{1}{\bar{\tau}}h^{(0)ab}\partial_{b}\omega\partial_{a}\sigma+\mathcal{O}\left(\frac{1}{\bar{\tau}^{2}}\right)~, \\
& h^{(0)}_{ab}(\phi^{c})= h^{(0)}_{ab}(\bar{\phi}^{c})+\frac{1}{\bar{\tau}}h^{(0)cd}\partial_{d}\omega\left(\partial_{c}h^{(0)}_{ab}\right)
\nonumber
\\
 & \qquad \qquad +\frac{1}{\bar{\tau}^{2}}\left[G^{(2)~c}\left(\partial_{c}h^{(0)}_{ab}\right)+\frac{1}{2}\left(h^{(0)cp}\partial_{p}\omega\right) \left(h^{(0)dq}\partial_{q}\omega\right)\partial_{c}\partial_{d}h^{(0)}_{ab}\right]+\mathcal{O}\left(\frac{1}{\bar{\tau}^{3}}\right)~,
\\
 & h^{(1)}_{ab}(\phi^{c}) = h^{(1)}_{ab}(\bar{\phi}^{c})+\frac{1}{\bar{\tau}}h^{(0)cd}\partial_{d}\omega\left(\partial_{c}h^{(1)}_{ab}\right)+\mathcal{O}\left(\frac{1}{\bar{\tau}^{2}}\right)~.
\end{align}
Inserting these expressions in  the full metric we can read the changes in the first order fields,
\begin{align}
& \sigma \rightarrow  \sigma \\
& h^{(1)}_{ab} \rightarrow  h^{(1)}_{ab}-2\omega h^{(0)}_{ab}+h^{(0)cd}\partial_{d}\omega\left(\partial_{c}h^{(0)}_{ab}\right)+h^{(0)}_{pb}\partial_{a}\left(h^{(0)pq}\partial_{q}\omega\right)+h^{(0)}_{pa}\partial_{b}\left(h^{(0)pq}\partial_{q}\omega\right).
\end{align}
This last expression can be more conveniently written as,
\be
 h^{(1)}_{ab} \rightarrow h^{(1)}_{ab}+2\mathcal{D}_{a} \mathcal{D}_{b}\omega-2\omega h^{(0)}_{ab}.\label{delta_omega_h1}
\ee
In order to preserve the original form of the metric, we must choose,
\begin{align}
&F^{(2)}=\sigma \omega+h^{(0)ab}\partial_{b}\omega\partial_{a}\sigma-\frac{1}{2}h^{(0)}_{ab}\left(h^{(0)ac}\partial_{c}\omega\right)\left(h^{(0)bd}\partial_{d}\omega\right),
\\
&2G^{(2)}_{a}=-\partial_{a}F^{(2)}+2\sigma\partial_{a}\omega-\left(\partial_{b}\omega\right)\partial_{a}\left(h^{(0)bp}\partial_{p}\omega\right)+2\omega \partial_{a}\omega-h^{(1)}_{ab}\partial^{b}\omega-\partial^{b}\omega \partial^{c}\omega\left(\partial_{c}h^{(0)}_{ab}\right).
 \end{align} 
This results in the transformation for  $h^{(2)}_{ab}$ as
\begin{align}
 h^{(2)}_{ab} \rightarrow 
 \label{h2ab}
&~h^{(2)}_{ab}-\mathcal{D}_{a}\omega \mathcal{D}_{b}\omega+\left[-\omega h^{(1)}_{ab}+ \mathcal{D}^{c}\omega \mathcal{D}_{c} h^{(1)}_{ab}+2h^{(1)c}_{(a}\mathcal{D}_{b)}\mathcal{D}_{c}\omega \right]+\left(2F^{(2)}+\omega^{2}\right)h^{(0)}_{ab}
\nonumber
\\
& +2\mathcal{D}_{(a}G^{(2)}_{b)}-4\omega~\mathcal{D}_{a}\mathcal{D}_{b}\omega+ \mathcal{D}_{a}\mathcal{D}^{c}\omega~\mathcal{D}_{b}\mathcal{D}_{c}\omega
+2\mathcal{D}_{(a}\mathcal{D}^{c}\omega~\mathcal{D}^{d}\omega~\Gamma^{(0)}_{b)cd}
\nonumber
\\
& +\mathcal{D}^{c}\omega~\mathcal{D}^{d}\omega \left(\mathcal{D}_{c}\Gamma^{(0)}_{(ba)d}+\Gamma^{(0)p}_{cd}\Gamma^{(0)}_{(ba)p}-\Gamma^{(0)p}_{(ac}\Gamma^{(0)}_{b)dp}\right),
\end{align}
where we have used the following notation
\be
\mathcal{D}_{c}\Gamma^{(0)}_{pqr} := \partial_{c}\Gamma^{(0)}_{pqr}-\Gamma^{(0)i}_{cp}\Gamma^{(0)}_{iqr}-\Gamma^{(0)i}_{cq}\Gamma^{(0)}_{pir}-\Gamma^{(0)i}_{cr}\Gamma^{(0)}_{pqi}~.
\ee
and the following results 
\begin{align}
& \Gamma^{(0)}_{qab}+\Gamma^{(0)}_{aqb}=\partial_{b}h^{(0)}_{qa},
\\
& \Gamma^{(0)}_{cad}+\Gamma^{(0)}_{dca} =\partial_{a}h^{(0)}_{cd},
\\
& \partial_{p}\omega~\partial_{a}h^{(0)cp}+h^{(0)cp}\partial_{a}\partial_{p}\omega 
=\mathcal{D}_{a}\mathcal{D}^{c}\omega-\mathcal{D}^{d}\omega~\Gamma^{c~(0)}_{ad}.
\end{align}
Next we  simplify \ref{h2ab}.
Defining $\phi_{a}= \mathcal{D}_{a} \phi$, for any scalar function $\phi$, we obtain,
\begin{align}
\delta_{\omega}h^{(2)}_{ab}
= & -\omega_{a}\omega_{b}-\omega k_{ab}+2\sigma \omega h^{(0)}_{ab}+\omega^{c} \mathcal{D}_{c}\,k_{ab}+\omega^{c}_{b}\,k_{ac}+\omega^{c}_{a}\,k_{bc}-2\sigma_{c}\omega^{c}h^{(0)}_{ab}-4\sigma \omega_{ab}
\nonumber
\\
&  +\left(2\sigma \omega+2\omega_{c}\sigma^{c}-\omega_{c}\omega^{c}+\omega^{2}\right)h^{(0)}_{ab}-4\omega \omega_{ab}+\omega^{c}_{a}\omega_{bc}+\omega^{c}_{a}\omega^{d}\Gamma^{(0)}_{bcd}+\omega^{c}_{b}\omega^{d}\Gamma^{(0)}_{acd}
\nonumber
\\
&  -\frac{1}{2}\left(\mathcal{D}_{a}\mathcal{D}_{b}+ \mathcal{D}_{b}\mathcal{D}_{a}\right)F^{(2)}+\sigma_{a}\omega_{b}+2\sigma \omega_{ab}+\sigma_{b}\omega_{a}
\nonumber
\\
&  -\frac{1}{2}\omega^{c}\left(\mathcal{D}_{a}\,k_{bc}\right)+\sigma_{a}\omega_{b}-\frac{1}{2}k_{bc}\omega^{c}_{a}+2\sigma \omega_{ab}-\frac{1}{2}\omega^{c}\left(\mathcal{D}_{b}\,k_{ac}\right)+\sigma_{b}\omega_{a}-\frac{1}{2}\,k_{ac}\omega^{c}_{b}
\nonumber
\\
&  -\omega_{ac}\omega^{c}_{b}-\frac{1}{2}\omega_{c}\left(\mathcal{D}_{a} \mathcal{D}_{b}+ \mathcal{D}_{b}\mathcal{D}_{a}\right)\omega^{c}+2\omega_{a}\omega_{b}+2\omega \omega_{ab}
\nonumber
\\
&  -\frac{1}{2}\omega^{p}\omega^{q}\left(\mathcal{D}_{a}\Gamma^{(0)}_{bpq}+ \mathcal{D}_{b}\Gamma^{(0)}_{apq} \right)-\omega^{p}_{a}\omega^{q}\Gamma^{(0)}_{bpq}-\omega^{p}_{b}\omega^{q}\Gamma^{(0)}_{apq}
\nonumber
\\
&  +\omega^{c}\omega^{d} \left(\mathcal{D}_{c}\Gamma^{(0)}_{(ba)d}+\Gamma^{(0)~p}_{cd}\,\Gamma^{(0)}_{(ba)p}-\Gamma^{(0)~p}_{(ac}\,\Gamma^{(0)}_{b)dp}\right).
\end{align}
We further obtain,
\begin{align}
\left(\mathcal{D}_{a}\mathcal{D}_{b}+ \mathcal{D}_{b}\mathcal{D}_{a}\right)F^{(2)}
= &~2\sigma_{ab}\omega+2\omega_{a}\sigma_{b}+2\sigma_{a}\omega_{b}+2\sigma \omega_{ab}
\nonumber
\\
& +\omega_{c}\left(\mathcal{D}_{a}\mathcal{D}_{b}+ \mathcal{D}_{b}\mathcal{D}_{a}\right)\sigma^{c}+2\sigma_{ca}\omega^{c}_{b}+2\sigma_{cb}\omega_{a}^{c}+\sigma_{c}\left(\mathcal{D}_{a}\mathcal{D}_{b}+ \mathcal{D}_{b}\mathcal{D}_{a}\right)\omega^{c}
\nonumber
\\
& -\omega_{ac}\omega^{c}_{b}-\omega_{bc}\omega^{c}_{a}-\omega_{c}\left(\mathcal{D}_{a}\mathcal{D}_{b}+ \mathcal{D}_{b}\mathcal{D}_{a}\right)\omega^{c}.
\end{align}
Combining these expressions we obtain,
\begin{align}
\delta_{\omega}h^{(2)}_{ab}
=& -\omega \,k_{ab}+\omega^{c} \mathcal{D}_{c}\,k_{ab}+\frac{1}{2}\omega^{c}_{b}k_{ac}+\frac{1}{2}\omega^{c}_{a}\,k_{bc}-\frac{1}{2}\omega^{c}\left(\mathcal{D}_{a}\,k_{bc}\right)-\frac{1}{2}\omega^{c}\left(\mathcal{D}_{b} \,k_{ac}\right)
\nonumber
\\
& +2\sigma \omega h^{(0)}_{ab}+\sigma_{(a}\omega_{b)}-\sigma \omega_{ab}-\sigma_{c(a}\omega^{c}_{b)}-\sigma_{c}\omega^{c}_{(ab)}+\left(\sigma \leftrightarrow \omega \right)
\nonumber
\\
& +\omega_{a}\omega_{b}+\left(-\omega_{c}\omega^{c}+\omega^{2}\right)h^{(0)}_{ab}-2\omega \omega_{ab}+\omega^{c}_{a}\omega_{bc}
\nonumber
\\
& +\frac{1}{2}\omega^{c}\omega^{d} \Big[\left(\mathcal{D}_{c}\Gamma^{(0)}_{bad}- \mathcal{D}_{a}\Gamma^{(0)}_{bcd}-\Gamma^{(0)\,p}_{ac}\Gamma^{(0)}_{bdp}+\Gamma^{(0)\,p}_{cd}\Gamma^{(0)}_{bap}\right)
\nonumber
\\
&\quad \quad \quad \quad +\left(\mathcal{D}_{c}\Gamma^{(0)}_{abd}- \mathcal{D}_{b}\Gamma^{(0)}_{acd}+\Gamma^{(0)\,p}_{cd}\,\Gamma^{(0)}_{abp}-\Gamma^{(0)\,p}_{bc}\,\Gamma^{(0)}_{adp}\right)\Big].
\end{align}
Using
\begin{align}
& \mathcal{D}_{c}\Gamma^{(0)}_{bad}- \mathcal{D}_{a} \Gamma^{(0)}_{bcd}-\Gamma^{(0)\,p}_{ac}\,\Gamma^{(0)}_{bdp}+\Gamma^{(0)\,p}_{cd}\,\Gamma^{(0)}_{bap} =R^{(0)}_{bdca}=-h^{(0)}_{bc}h^{(0)}_{da}+h^{(0)}_{ba}h^{(0)}_{dc}, \\
& \mathcal{D}_{c}\Gamma^{(0)}_{abd}- \mathcal{D}_{b}\Gamma^{(0)}_{acd}+\Gamma^{(0)\,p}_{cd}\,\Gamma^{(0)}_{abp}-\Gamma^{(0)\,p}_{bc}\,\Gamma^{(0)}_{adp}=R^{(0)}_{adcb}=-h^{(0)}_{ac}h^{(0)}_{bd}+h^{(0)}_{ab}h^{(0)}_{cd},
\end{align}
we obtain our final form for $\delta_{\omega}h^{(2)}_{ab}$,
\begin{align}
\delta_{\omega}h^{(2)}_{ab}
=& -\omega\,k_{ab}+\omega^{c} \mathcal{D}_{c}\,k_{ab}+\frac{1}{2}\omega^{c}_{b}\,k_{ac}+\frac{1}{2}\omega^{c}_{a}\,k_{bc}-\frac{1}{2}\omega^{c}\left(\mathcal{D}_{a}\,k_{bc}\right)-\frac{1}{2}\omega^{c}\left(\mathcal{D}_{b}\,k_{ac}\right)
\nonumber
\\
&  +2\sigma \omega h^{(0)}_{ab}+\sigma_{(a}\omega_{b)}-\sigma \omega_{ab}-\sigma_{c(a}\omega^{c}_{b)}-\sigma_{c}\omega^{c}{}_{(ab)}+\left(\sigma \leftrightarrow \omega \right)
\nonumber
\\
&  +\omega^{2}h^{(0)}_{ab}-2\omega \omega_{ab}+\omega^{c}_{a}\omega_{bc}. \label{delta_omega_h2}
\end{align}


\section{Expansion of the equations of motion}
\label{appendix_EOM}

Given the previous series of coordinate transformations, we arrive at the following form of  the asymptotic metric, near timelike infinity,
\be
\label{Beig_Schmidt_form_APP}
ds^{2}=-N^{2}d\tau^{2}+h_{ab}d\phi^{a}d\phi^{b}, 
\ee
where
\bea
&& N  =1+\frac{\sigma(\phi^{a})}{\tau}, \\
&& h_{ab}=\tau^{2}\left[h^{(0)}_{ab}(\phi^{c})+\frac{1}{\tau}h^{(1)}_{ab}(\phi^{c})+\frac{1}{\tau^{2}}h^{(2)}_{ab}(\phi^{c})+\mathcal{O}\left(\frac{1}{\tau^{3}}\right)\right].
\eea
The future directed unit normal vector to a $\tau=\textrm{constant}$ surface is,
\be
n_{\mu}=-N\nabla_{\mu}\tau, \qquad n^{\mu}=\frac{1}{N}\delta^{\mu}_{\tau}~. \label{unit_normal}
\ee
The induced metric on $\tau=\textrm{constant}$ hypersurface is $h_{ab}$, while the inverse spatial metric has the following expansion,
\begin{align}
h^{ab}=\frac{1}{\tau^{2}}h^{(0)ab}-\frac{1}{\tau^{3}}h^{(1)ab}-\frac{1}{\tau^{4}}\left(h^{(2)ab}-h^{(1)a}_{c}h^{(1)cb}\right)+\mathcal{O}\left(\frac{1}{\tau^{5}}\right)~.
\end{align}
For any spatial tensor $T^{(n)}_{ab}$ at order $n$ in the expansion, we raise and lower indices with $h^{(0)}_{ab}$, for example, 
\be
T^{(n)ab}=h^{(0)ac}h^{(0)bd}T^{(n)}_{cd}.
\ee 
For a general spatial tensor $T_{ab}$, we have $T^{ab}=h^{ac}h^{bd}T_{cd}$.

\subsection*{The extrinsic curvature $K_{ab}$}

The extrinsic curvature of  $\tau=\textrm{constant}$ hypersurface takes the form,
\begin{align}
 K_{ab}= \frac{1}{2N}\partial_{\tau}h_{ab}
& =\tau h^{(0)}_{ab}+\left(\frac{1}{2}h^{(1)}_{ab}-\sigma h^{(0)}_{ab}\right)+\frac{1}{\tau}\left(\sigma^{2}h^{(0)}_{ab}-\frac{\sigma}{2}h^{(1)}_{ab} \right)+\mathcal{O}\left(\frac{1}{\tau^{2}}\right).
\end{align}
Upon raising one and two indices respectively  we have
\begin{align}
K^{a}_{b}=h^{ac}K_{cb}=&\frac{1}{\tau}\delta^{a}_{b}+\frac{1}{\tau^{2}}\left(-\frac{1}{2}h^{(1)a}_{b}-\sigma \delta^{a}_{b} \right)
\nonumber
\\
&
+\frac{1}{\tau^{3}}\left(\sigma^{2}\delta^{a}_{b}+\frac{\sigma}{2}h^{(1)a}_{b}-h^{(2)a}_{b}+\frac{1}{2}h^{(1)ap}h^{(1)}_{pb} \right)+\mathcal{O}\left(\frac{1}{\tau^{4}}\right)
\\
K^{ab} = h^{ac}K^{b}_{c} =&\frac{1}{\tau^{3}}h^{(0)ab}+\frac{1}{\tau^{4}}\left(-\frac{3}{2}h^{(1)ab}-\sigma h^{(0)ab}\right)
\nonumber
\\
& +\frac{1}{\tau^{5}}\left(-2h^{(2)ab}+2h^{(1)ap}h^{(1)b}_{p}+\frac{3\sigma}{2}h^{(1)ab}+\sigma^{2}h^{(0)}_{ab}\right)+\mathcal{O}\left(\frac{1}{\tau^{6}}\right)~.
\end{align}
The trace of the extrinsic curvature becomes,
\begin{align}
K&=\delta^{b}_{a}K^{a}_{b}=\frac{3}{\tau}+\frac{1}{\tau^{2}}\left(-\frac{1}{2}h^{(1)}-3\sigma\right)+\frac{1}{\tau^{3}}\left(3\sigma^{2}+\frac{\sigma}{2}h^{(1)}-h^{(2)}+\frac{1}{2}h^{(1)ab}h^{(1)}_{ab} \right)+\mathcal{O}\left(\frac{1}{\tau^{4}}\right)~.
\end{align}
 
\subsection*{Asymptotic expansion of intrinsic geometry}

For any perturbed symmetric, spatial tensor $S^{(n)}_{ab}$, we note the following identity
\begin{align}
-\partial_{d}S^{(n)}_{bc}+\partial_{b}S^{(n)}_{dc}+\partial_{c}S^{(n)}_{bd}
&=-\mathcal{D}_{d}S^{(n)}_{bc}+\mathcal{D}_{b}S^{(n)}_{dc}+ \mathcal{D}_{c}S^{(n)}_{bd}+2\Gamma^{(0)p}_{bc}S^{(n)}_{pd},
\end{align}
where $\mathcal{D}$ denotes covariant derivative compatible with $h^{(0)}_{ab}$ on $\cal{H}$.
Using the above identity, the asymptotic expansion of the Christoffel symbol takes the  form,
\bea
\Gamma^{a}_{bc} &=&\frac{1}{2}h^{ad}\left(-\partial_{d}h_{bc}+\partial_{b}h_{dc}+\partial_{c}h_{bd}\right) \\
&\equiv& \Gamma^{(0)a}_{bc}+\frac{1}{\tau}\Gamma^{(1)a}_{bc}+\frac{1}{\tau^{2}}\Gamma^{(2)a}_{bc}+\mathcal{O}\left(\frac{1}{\tau^{3}}\right)~,
\eea
where
\begin{align}
\Gamma^{(1)a}_{bc} &= -h^{(1)a}_{d}\Gamma^{(0)d}_{bc}+\frac{1}{2}h^{(0)ad}\left(-\mathcal{D}_{d}h^{(1)}_{bc}+ \mathcal{D}_{b}h^{(1)}_{dc}+\mathcal{D}_{c}h^{(1)}_{bd}+2\Gamma^{(0)p}_{bc}h^{(1)}_{pd}\right), \\
\Gamma^{(2)a}_{bc} &= -\left(h^{(2)a}_{d}-h^{(1)a}_{p}h^{(1)p}_{d}\right)\Gamma^{(0)d}_{bc} +\frac{1}{2}h^{(0)ad}\left(-\mathcal{D}_{d}h^{(2)}_{bc}+ \mathcal{D}_{b}h^{(2)}_{dc}+ \mathcal{D}_{c}h^{(2)}_{bd}+2\Gamma^{(0)p}_{bc}h^{(2)}_{pd}\right) \nonumber \\
&\hskip 2 cm -\frac{1}{2}h^{(1)ad}\left(-\mathcal{D}_{d}h^{(1)}_{bc}+ \mathcal{D}_{b}h^{(1)}_{dc}+ \mathcal{D}_{c}h^{(1)}_{bd}+2\Gamma^{(0)p}_{bc}h^{(1)}_{pd}\right).
\end{align}
Here, $\Gamma^{(0)a}_{bc}$ is the non-tensorial Christoffel symbol associated with the zeroth order spatial metric $h^{(0)}_{ab}$. The other expansion coefficients are tensors and have the following simplified expressions,
\begin{align}
\Gamma^{(1)a}_{bc}&=\frac{1}{2}\left(-\mathcal{D}^{a}h^{(1)}_{bc}+\mathcal{D}_{b}h^{(1)a}_{c}+ \mathcal{D}_{c}h^{(1)a}_{b}\right)~,
\\
\Gamma^{(2)a}_{bc}
&=\frac{1}{2}\left(-\mathcal{D}^{a}h^{(2)}_{bc}+ \mathcal{D}_{b}h^{(2)a}_{c}+ \mathcal{D}_{c}h^{(2)a}_{b}\right)-\frac{1}{2}h^{(1)ad}\left(-\mathcal{D}_{d}h^{(1)}_{bc}+ \mathcal{D}_{b}h^{(1)}_{dc}+\mathcal{D}_{c}h^{(1)}_{bd}\right)~.
\end{align}
The three-dimensional Ricci tensor takes the  form,
\begin{align}
\mathcal{R}_{ab}&= \partial_{c}\Gamma^{c}_{ab}-\partial_{b}\Gamma^{c}_{ca}+\Gamma^{c}_{ab}\Gamma^{d}_{cd}-\Gamma^{c}_{ad}\Gamma^{d}_{bc}
\nonumber
\\
 &\equiv \mathcal{R}^{(0)}_{ab}+\frac{1}{\tau}\mathcal{R}^{(1)}_{ab}+\frac{1}{\tau^{2}}\mathcal{R}^{(2)}_{ab} +\mathcal{O}\left(\frac{1}{\tau^{3}}\right).
\end{align}
Here, $\mathcal{R}^{(0)}_{ab}$ is the Ricci tensor associated with the spatial metric $h^{(0)}_{ab}$, while the other two expansion coefficients are,
\begin{align}
\mathcal{R}^{(1)}_{ab}
 =&~\frac{1}{2}\left(\mathcal{D}_{c}\mathcal{D}_{a}h^{(1)c}_{b}+\mathcal{D}_{c}\mathcal{D}_{b}h^{(1)c}_{a}- \mathcal{D}_{c}\mathcal{D}^{c}h^{(1)}_{ab}-\mathcal{D}_{a}\mathcal{D}_{b}h^{(1)} \right)~,
\\
\mathcal{R}^{(2)}_{ab}
= &~\frac{1}{2}\left(\mathcal{D}_{c}\mathcal{D}_{a}h^{(2)c}_{b}+ \mathcal{D}_{c}\mathcal{D}_{b}h^{(2)c}_{a}- \mathcal{D}_{c}\mathcal{D}^{c}h^{(2)}_{ab}-\mathcal{D}_{a}\mathcal{D}_{b}h^{(2)} \right)
\nonumber
\\
&  +\frac{1}{2}\mathcal{D}_{b}\left(h^{(1)cd}\mathcal{D}_{a}h^{(1)}_{cd} \right)
-\frac{1}{2}\mathcal{D}_{c}\left[h^{(1)cd}\left(-\mathcal{D}_{d}h^{(1)}_{ab}+ \mathcal{D}_{b}h^{(1)}_{da}+\mathcal{D}_{a}h^{(1)}_{bd}\right)\right]
\nonumber
\\
&  +\frac{1}{4}\mathcal{D}_{c}h^{(1)}\left(-\mathcal{D}^{c}h^{(1)}_{ba}+ \mathcal{D}_{b}h^{(1)c}_{a}+ \mathcal{D}_{a}h^{(1)c}_{b}\right)
\nonumber
\\
& -\frac{1}{4}\mathcal{D}_{a}h^{(1)c}_{d}\mathcal{D}_{b}h^{(1)d}_{c}+\frac{1}{2}\mathcal{D}^{c}h^{(1)}_{ad}\mathcal{D}_{c}h^{(1)d}_{b}-\frac{1}{2}\mathcal{D}^{c}h^{(1)}_{ad}\mathcal{D}^{d}h^{(1)}_{bc}~.
\end{align}
These expressions will be used extensively in what follows.

\subsection*{The Hamiltonian constraint}

The Hamiltonian constraint takes the form,
\begin{align}
H \equiv \frac{1}{N}\partial_{\tau}K+K_{ab}K^{ab}-\frac{1}{N}h^{ab}D_{a}D_{b}N =0,
\end{align}
where $D_{a}$ is the  covariant derivative compatible with $h_{ab}$. Expanding out each of these terms we obtain,
\begin{align}
H 
&= \frac{H^{(0)}}{\tau^{2}}+\frac{H^{(1)}}{\tau^{3}}+\frac{H^{(2)}}{\tau^{4}}+\mathcal{O}\left(\frac{1}{\tau^{5}}\right)~.
\end{align}
where
\begin{align}
&\quad H^{(0)}=0,
\\
&\quad H^{(1)}=\left(-\square+3\right)\sigma =0,
\\
&\quad H^{(2)}=h^{(2)}-9\sigma^{2}-\frac{1}{4}h^{(1)ab}h^{(1)}_{ab}-\frac{1}{2}\sigma h^{(1)}+h^{(1)ab}\mathcal{D}_{a}\mathcal{D}_{b}\sigma+\sigma \square \sigma+h^{(0)ab}\Gamma^{(1)c}_{ab}\mathcal{D}_{c}\sigma =0.
\end{align}
 Using,  $k_{ab}= h^{(1)}_{ab}+2\sigma h^{(0)}_{ab}$, cf.~\ref{def_kab}, the second order  coefficient  can be simplified, yielding,
\begin{align}
H^{(2)}=&~h^{(2)}-12\sigma^{2}-\frac{1}{4}k^{ab}k_{ab}+k^{ab}\mathcal{D}_{a}\mathcal{D}_{b}\sigma+ \mathcal{D}_{c}\sigma \mathcal{D}^{c}\sigma
\nonumber
\\
& +\frac{1}{2}\sigma k -\sigma \left(\square-3\right)\sigma-\frac{1}{2} \mathcal{D}_{c}\sigma \left(\mathcal{D}^{c}k \right)+ \mathcal{D}_{c}\sigma \mathcal{D}_{a}k^{ac}~.
\end{align}
Now upon using our boundary condition $k=0$ and lower order equations of motion it simplifies to 
\begin{align}
H^{(2)}&=h^{(2)}-12\sigma^{2}-\frac{1}{4}k^{ab}k_{ab}+k^{ab}\mathcal{D}_{a}\mathcal{D}_{b}\sigma+ \mathcal{D}_{c}\sigma \mathcal{D}^{c}\sigma = 0.
\label{H2_APP}
\end{align}


\subsection*{The momentum constraint}

The   momentum constraint $H_{a}=0$ takes the form,
\begin{align}
H_{a} \equiv D_{b}K^{b}_{a}-D_{a}K =0.
\end{align}
This can be expanded as,
\begin{align}
H_{a}
&= \frac{1}{\tau}H_{a}^{(0)}+\frac{1}{\tau^{2}}H_{a}^{(1)}+\frac{1}{\tau^{3}}H_{a}^{(2)} + \ldots
\end{align}
where
\begin{align}
H_{a}^{(0)}=&0\\
H_{a}^{(1)}
=&-\frac{1}{2}\mathcal{D}_{b}\left(k^{b}_{a} - k \delta^{b}_{a} \right)\\
H_{a}^{(2)}
 = &-\mathcal{D}_{b}h^{(2)b}_{a}+\frac{1}{2}k^{bp}\left(\mathcal{D}_{b}k_{pa}\right)+\frac{1}{2}k_{pa}\left(\mathcal{D}_{b}k^{bp}\right)-\frac{3}{2}\sigma \left(\mathcal{D}_{b}k^{b}_{a}\right)-\frac{1}{4}k^{c}_{a}\left(\mathcal{D}_{c}k\right)+\frac{\sigma}{2}\left(\mathcal{D}_{a}k\right)
\nonumber
\\
&+ \mathcal{D}_{a}\left[h^{(2)}-\frac{3}{8}k^{bc}k_{bc}+\sigma k -4\sigma^{2}\right]~.
\end{align}
Using  second order Hamiltonian constraint $H^{(2)}=0$ and boundary condition $k=0$, together with first order equations of motion, we get
\begin{align} \label{H2a_APP}
H_{a}^{(2)}&=-\mathcal{D}_{b}h^{(2)b}_{a}+\frac{1}{2}k^{bp}\left(\mathcal{D}_{b}k_{pa}\right)
+\mathcal{D}_{a}\left(-\frac{1}{8}k^{bc} k_{bc}+8\sigma^{2}-k^{ab}\mathcal{D}_{a}\mathcal{D}_{b}\sigma- \mathcal{D}_{c}\sigma \mathcal{D}^{c}\sigma\right)=0.
\end{align}
\subsection*{Asymptotic expansion of the evolution equation }

The evolution equation of the spatial metric $h_{ab}$ takes the following form,
\begin{align}
H_{ab}:=\mathcal{R}_{ab}+\frac{1}{N}\partial_{\tau}K_{ab}-2K_{ac}K^{c}_{b}+KK_{ab}-\frac{1}{N}D_{a}D_{b}N =0.
\end{align}
Expanding in powers of $\frac{1}{\tau}$ we have,
\begin{align}
H_{ab}
&\equiv H^{(0)}_{ab}+\frac{1}{\tau}H^{(1)}_{ab}+\frac{1}{\tau^{2}}H^{(2)}_{ab} + \ldots,
\end{align}
where
\begin{align}
H^{(0)}_{ab}&=\mathcal{R}^{(0)}_{ab}+2h^{(0)}_{ab} =0, \\
H^{(1)}_{ab}&=-\frac{1}{2}\left(\square +3 \right) k_{ab} =0, \\
H_{ab}^{(2)}&=-\frac{1}{2}\left(\square +2\right)h^{(2)}_{ab}+T^{(kk)}_{ab}+T^{(k\sigma)}_{ab}+T^{(\sigma \sigma)}_{ab} =0, \label{H2ab}
\end{align}
and where the non-linear terms are 
\begin{align}
&\quad T^{(kk)}_{ab} =\frac{1}{2}\Bigg[\left(\mathcal{D}_{c} k_{d(a}\mathcal{D}_{b)}k^{cd}\right)-\frac{1}{2}\mathcal{D}_{a}k^{cd}\mathcal{D}_{b}k_{cd}
+\left(\mathcal{D}^{c} k_{ad}\right)\left(\mathcal{D}_{c} k^{d}_{b}\right)-\left(\mathcal{D}^{c} k_{ad}\right)\left(\mathcal{D}^{d} k_{bc}\right)
\nonumber
\\
&\hskip 2 cm  -k^{p}_{a} k_{pb}+k^{cd}\left(\mathcal{D}_{c}\mathcal{D}_{d} k_{ab}-\mathcal{D}_{c}\mathcal{D}_{(a} k_{b)d}\right)\Bigg] , \\
&\quad   T^{(k\sigma)}_{ab} =\frac{1}{2}\Bigg[-\mathcal{D}_{a}\mathcal{D}_{b}\left(k^{cd}\mathcal{D}_{c}\mathcal{D}_{d}\sigma\right)+4\mathcal{D}^{c}\sigma \left(-\mathcal{D}_{c} k_{ab}+ \mathcal{D}_{(a} k_{b)c}\right)-4\sigma k_{ab}
\nonumber
\\
&\hskip 2 cm +  \left(-2h^{(0)}_{ab}k^{cd}\mathcal{D}_{c}\mathcal{D}_{d}\sigma +4k^{cd}h^{(0)}_{d(a}\mathcal{D}_{b)}\mathcal{D}_{c}\sigma\right)\Bigg]
, \\
&\quad  T^{(\sigma \sigma)}_{ab} =\frac{1}{2}\Bigg[\mathcal{D}_{a}\mathcal{D}_{b}\left(5\sigma^{2}- \mathcal{D}_{c}\sigma \mathcal{D}^{c}\sigma\right)
+h^{(0)}_{ab}\left(18\sigma^{2}+4\mathcal{D}^{c}\sigma \mathcal{D}_{c}\sigma\right)
+4\sigma \mathcal{D}_{a}\mathcal{D}_{b}\sigma\Bigg],
\end{align}
where we have used the boundary condition $k=0$ and the first order equations of motion.

\section{A consistency check}
\label{a_consistency_check}

 In this appendix we perform a non-trivial consistency check on our asymptotic equations of motion and  expression \ref{delta_omega_h2} for $\delta_\omega h^{(2)}_{ab}$. We consider doing  a supertranslation on flat spacetime. Thus to begin with we have (for flat spacetime)
\be
\sigma = 0, \quad h^{(1)}_{ab} =0, \quad h^{(2)}_{ab} =0.
\ee
We note that $\sigma=0$ does not change under supertranslations. Thus for the supertranslated spacetime too  $\sigma = 0$ and  from \ref{delta_omega_h1} it follows that 
\be
h^{(1)}_{ab}= \delta_{\omega}h^{(1)}_{ab}=k_{ab}=-2\omega h^{(0)}_{ab}+2\omega_{ab}. \label{h1_flat_omega}
\ee
From  \ref{delta_omega_h2} it follows that 
\begin{align}
h^{(2)}_{ab}
=  \omega^{2}h^{(0)}_{ab}-2\omega \omega_{ab}+\omega^{c}_{a}\omega_{bc}. \label{h2_flat_omega}
\end{align}
We check that expression  \ref{h1_flat_omega} for $k_{ab}$ and \ref{h2_flat_omega} for $h^{(2)}_{ab}$  are consistent with second order equations of motion.

Recall that $\square \omega=3\omega$, and also we note the following useful relation,
\begin{align}
\square \omega_{a}&=\mathcal{D}_{b}\mathcal{D}^{b}\mathcal{D}_{a}\omega=\mathcal{D}_{b}\mathcal{D}_{a}\mathcal{D}^{b}\omega=[\mathcal{D}_{b},\mathcal{D}_{a}]\mathcal{D}^{b}\omega+\mathcal{D}_{a}\square \omega
\nonumber
\\
&=R^{(0)b}_{\,\,cba}\mathcal{D}^{c}\omega+3\omega_{a}=-2h^{(0)}_{ca}\omega^{c}+3\omega_{a}=\omega_{a}.
\end{align}

\subsection*{Hamiltonian constraint} 
 The Hamiltonian constraint \ref{H2_APP} becomes,
\begin{align}
h^{(2)}=\frac{1}{4}k_{ab}k^{ab}.
\end{align}
Given  expression \ref{h1_flat_omega} for $k_{ab}$, we have
\begin{align}
\label{hamiltonian_supertranslation}
\frac{1}{4}k_{ab}k^{ab}=3\omega^{2}-2\omega \square \omega+\omega_{ab}\omega^{ab}=-3\omega^{2}+\omega_{ab}\omega^{ab},
\end{align}
which matches with the trace of \ref{h2_flat_omega}, viz., 
\begin{align}
h^{(2)}&=3\omega^{2}-2\omega \square \omega+\omega^{ab}\omega_{ab}=-3\omega^{2}+\omega^{ab}\omega_{ab}.
\end{align}

\subsection*{Momentum constraint} 

The momentum constraint  presented in  \ref{H2a_APP} becomes,
\begin{align}\label{momentum_supertranslation}
\mathcal{D}^{b}h^{(2)}_{ab}&=\frac{1}{2}k^{bp}\,\mathcal{D}_{b}k_{pa} -\frac{1}{4}k^{bc}\,\mathcal{D}_{a} k_{bc}.
\end{align}
On the one hand, the right hand side  of  \ref{momentum_supertranslation} is
\begin{align}
\frac{1}{2}k^{bp}\,\mathcal{D}_{b}k_{pa} -\frac{1}{4}k^{bc}\,\mathcal{D}_{a} k_{bc}
&=-2\omega_{b}\omega^{b}_{a}+\omega^{bc}\mathcal{D}_{b}\mathcal{D}_{c}\omega_{a}+3\omega \omega_{a}-\omega^{bc}R^{(0)}_{cdab}\omega^{d}
\nonumber
\\
&=-2\omega_{b}\omega^{b}_{a}+\omega^{bc}\mathcal{D}_{b}\mathcal{D}_{c}\omega_{a}+3\omega \omega_{a}-\omega^{bc}\left(-h^{(0)}_{ac}h^{(0)}_{bd}+h^{(0)}_{bc}h^{(0)}_{ad}\right)\omega^{d}
\nonumber
\\
&=-\omega_{b}\omega^{b}_{a}+\omega^{bc}\mathcal{D}_{b}\mathcal{D}_{c}\omega_{a}.
\label{momentum_supertranslation2}
\end{align}
 On the other hand, the divergence of  \ref{h2_flat_omega} yields for the left hand side of  \ref{momentum_supertranslation}
\begin{align}
\mathcal{D}^{b}h^{(2)}_{ab}&= \mathcal{D}^{b}\left(\omega^{2}h^{(0)}_{ab}-2\omega \omega_{ab}+\omega^{c}_{a}\omega_{cb}\right)
\nonumber
\\
&=2\omega \omega_{a}-2\omega^{b}\omega_{ab}-2\omega \square \omega_{a}+\omega_{a}^{\,\,cb}\omega_{cb}+\omega^{c}_{a}\square \omega_{c}
\nonumber
\\
&=-\omega^{b}\omega_{ab}+\omega_{a}^{\,\,cb}\omega_{cb},
\end{align}
which  matches with \ref{momentum_supertranslation2}. 

\subsection*{Evolution equation} 

The evolution equation  as presented in \ref{H2ab} is decomposed into several terms,
\begin{align}\label{evo_second}
\left(\square+2\right)h^{(2)}_{ab}&=\frac{1}{2}\left(\underbrace{\mathcal{D}_{c}\,k_{da}\mathcal{D}_{b}\,k^{cd}}_{\rm Term~1}+\underbrace{\mathcal{D}_{c}k^{(1)}_{db}\mathcal{D}_{a}\,k^{cd}}_{\rm Term~2}\right)
\underbrace{-\frac{1}{2}\mathcal{D}_{a}\,k^{cd}\mathcal{D}_{b}\,k_{cd}}_{\rm Term~3}
\nonumber
\\
&\hskip 1 cm +\underbrace{\left(\mathcal{D}^{c}\,k_{ad}\right)\left(\mathcal{D}_{c}\,k^{d}_{b}\right)}_{\rm Term~4}-\underbrace{\left(\mathcal{D}^{c}\,k_{ad}\right)\left(\mathcal{D}^{d}\,k_{bc}\right)}_{\rm Term~5}-\underbrace{k^{p}_{a}\,k_{pb}}_{\rm Term~6}
\nonumber
\\
&\hskip 1 cm +\left(\underbrace{k^{cd}\mathcal{D}_{c}\mathcal{D}_{d}\,k_{ab}}_{\rm Term~7}\right)-\left(\frac{1}{2}\underbrace{k^{cd}\mathcal{D}_{c}\mathcal{D}_{a}\,k_{bd}}_{\rm Term~8}+\frac{1}{2}\underbrace{k^{cd}\mathcal{D}_{c}\mathcal{D}_{b}\,k_{ad}}_{\rm Term~9}\right)~.
\end{align}
We first evaluate the right hand side using   \ref{h1_flat_omega} and then evaluate the left hand side using \ref{h2_flat_omega} and show the match.

We obtain the following expression for various terms on the right hand side. For ``Term 1" we have,
\begin{align}
\frac{1}{2}\mathcal{D}_{c}\,k_{da}\mathcal{D}_{b}k^{cd}&=2\mathcal{D}_{c}\left(-\omega h_{da}^{(0)}+\omega_{da}\right)\mathcal{D}_{b}\left(-\omega h^{(0)cd}+\omega^{cd}\right)
\nonumber
\\
&=-2\omega^{c}[\mathcal{D}_{b},\mathcal{D}_{c}]\omega_{a}-2\omega^{c}\omega_{abc}+2\omega_{adc}[\mathcal{D}_{b},\mathcal{D}^{c}]\omega^{d}+2\omega_{adc}\omega_{b}^{\,\,dc}
\nonumber
\\
&=2h^{(0)}_{ab}\left(\omega_{c}\omega^{c}\right)-4\omega^{c}\omega_{abc}+2\omega_{adc}\omega_{b}^{\,\,dc}~,
\end{align}
while ``Term 2" follows from interchange of $(a,b)$ in ``Term 1''. For ``Term 3" we have, 
\begin{align}
\frac{1}{2}\mathcal{D}_{a}k^{cd}\mathcal{D}_{b}\,k_{cd}&=2\mathcal{D}_{a}\left(-\omega h^{(0)cd}+\omega^{cd}\right)\mathcal{D}_{b}\left(-\omega h^{(0)}_{cd}+\omega_{cd}\right)
\nonumber
\\
&=-6\omega_{a}\omega_{b}+ 2\left\{[\mathcal{D}_{a},\mathcal{D}^{c}]\omega^{d}+\omega_{a}^{\,\,dc}\right\}\left\{[\mathcal{D}_{b},\mathcal{D}_{c}]\omega_{d}+\omega_{bdc}\right\}
\nonumber
\\
&=2\omega_{c}\omega^{c}h^{(0)}_{ab}-4\omega^{c}\omega_{abc}+2\omega_{adc}\omega_{b}^{\,\,dc}~.
\end{align}
For ``Term 4" we have, 
\begin{align}
\left( \mathcal{D}^{c} k_{ad}\right)\left(\mathcal{D}_{c}k^{d}_{b}\right)&=4\mathcal{D}^{c}\left(-\omega h^{(0)}_{ad}+\omega_{ad}\right)D^{(0)}_{c}\left(-\omega \delta^{d}_{b}+\omega^{\,\,d}_{b}\right)
\nonumber
\\
&=4\omega_{c}\omega^{c}h^{(0)}_{ab}-8\omega^{c}\omega_{abc}+4\omega_{adc}\omega_{b}^{\,\,dc}.
\end{align}
For ``Term 5" we have, 
\begin{align}
\left(\mathcal{D}^{c}\,k_{ad}\right)\left(\mathcal{D}^{d}\,k_{bc}\right)&=4\mathcal{D}^{c}\left(-\omega h^{(0)}_{ad}+\omega_{ad}\right)\mathcal{D}^{d}\left(-\omega h^{(0)}_{bc}+\omega_{bc}\right)
\nonumber
\\
&=4\omega_{a}\omega_{b}-4\omega^{c}[\mathcal{D}_{b},\mathcal{D}_{c}]\omega_{a}-8\omega^{c}\omega_{abc}-4\omega^{c}[\mathcal{D}_{a},\mathcal{D}_{c}]\omega_{b}+4[\mathcal{D}_{c},\mathcal{D}_{d}]\omega_{a}\omega_{b}^{\,\,cd}+4\omega_{acd}\omega_{b}^{\,\,cd}
\nonumber
\\
&=-4\omega_{a}\omega_{b}+8\omega_{c}\omega^{c}h^{(0)}_{ab}-12\omega^{c}\omega_{abc}+4\omega_{acd}\omega_{b}^{\,\,cd}+4\omega^{c}\omega_{bca}
\nonumber
\\
&=-4\omega_{a}\omega_{b}+8\omega_{c}\omega^{c}h^{(0)}_{ab}-8\omega^{c}\omega_{abc}+4\omega_{acd}\omega_{b}^{\,\,cd}+4\omega^{c}[\mathcal{D}_{a},\mathcal{D}_{c}]\omega_{b}
\nonumber
\\
&=4\omega_{c}\omega^{c}h^{(0)}_{ab}-8\omega^{c}\omega_{abc}+4\omega_{acd}\omega_{b}^{\,\,cd}~.
\end{align}
For ``Term 6" we have, 
\begin{align}
  k^{p}_{a}\,k_{pb}&=4\left(-\omega \delta^{p}_{a}+\omega^{p}_{a}\right)\left(-\omega h^{(0)}_{pb}+\omega_{pb}\right)
\nonumber
\\
&=4\omega^{2}h^{(0)}_{ab}-8\omega \omega_{ab}+4\omega^{p}_{a}\omega_{pb}~.
\end{align}
For ``Term 7" we have,
\begin{align}
k^{cd}\,\mathcal{D}_{c}\, \mathcal{D}_{d}\,k_{ab}&=4\left(-\omega h^{(0)cd}+\omega^{cd}\right)\mathcal{D}_{c}\mathcal{D}_{d}\left(-\omega h^{(0)}_{ab}+\omega_{ab}\right)
\nonumber
\\
&=12\omega^{2} h^{(0)}_{ab}-4\omega^{cd}\omega_{cd}h^{(0)}_{ab}-4\omega \square \omega_{ab} +4\omega^{cd}\omega_{abdc}~,
\end{align}
and finally for ``Term 8" we have,
\begin{align}
\frac{1}{2}k^{cd}\,\mathcal{D}_{c}\,\mathcal{D}_{a}\, k_{bd}&=2\left(-\omega h^{(0)cd}+\omega^{cd}\right)\mathcal{D}_{c}\, \mathcal{D}_{a}\left(-\omega h^{(0)}_{db}+\omega_{bd}\right)
\nonumber
\\
&\hskip -2 cm =2\omega \omega_{ab}-2\omega_{ac}\omega^{c}_{b}-2\omega h^{(0)cd}\mathcal{D}_{c}\left\{[\mathcal{D}_{a},\mathcal{D}_{d}]\omega_{b} \right\}-2\omega \square \omega_{ab}+2\omega^{cd}\omega_{badc}+2\omega^{cd}\mathcal{D}_{c} \left\{[\mathcal{D}_{a},\mathcal{D}_{d}]\omega_{b} \right\}
\nonumber
\\
&\hskip -2 cm =6\omega^{2}h^{(0)}_{ab}-2\omega \square \omega_{ab}+2\omega^{cd}\omega_{badc}-2\omega_{cd}\omega^{cd}h^{(0)}_{ab}~.
\end{align}
``Term 9'' is obtained by interchanging $(a,b)$ in ``Term 8''. Collecting all these expressions we get,
\begin{align}\label{evo_second_super}
\left(\square+2\right)h^{(2)}_{ab}=-4\omega^{2}h^{(0)}_{ab}+8\omega \omega_{ab}+2h^{(0)}_{ab}\left(\omega_{c}\omega^{c}\right)
-4\omega^{c}_{a}\omega_{cb}-4\omega^{c}\omega_{abc}+2\omega_{adc}\omega_{b}^{\,\,dc}
\end{align}
On the other hand, using expression \ref{h2_flat_omega} for $h^{(2)}_{ab}$ we obtain,
\begin{align}
\left(\square+2\right)h^{(2)}_{ab}=~& 2\omega^{2}h^{(0)}_{ab}-4\omega \omega_{ab}+2\omega^{c}_{a}\omega_{cb}+h^{(0)}_{ab} \mathcal{D}^{c}\, \mathcal{D}_{c}\omega^{2}- 2\mathcal{D}^{c} \, \mathcal{D}_{c}\left(\omega \omega_{ab}\right)+ \mathcal{D}^{c} \, \mathcal{D}_{c}\left(\omega^{d}_{a}\omega_{db}\right)
\nonumber
\\
=~& 8\omega^{2}h^{(0)}_{ab}-10\omega \omega_{ab}+2h^{(0)}_{ab}\left(\omega_{c}\omega^{c}\right)+2\omega^{c}_{a}\omega_{cb}-4\omega^{c}\omega_{abc}+2\omega_{acd}\omega_{b}^{\,\,cd}
\nonumber
\\
& +\omega_{cb}\square\omega^{c}_{a}+\omega^{c}_{a}\square\omega_{cb}-2\omega\square\omega_{ab}~.
\end{align}
We now have the following identity,
\begin{align}
\square \omega_{ab}&= \mathcal{D}^{c}\, \mathcal{D}_{c}\, \mathcal{D}_{a}\, \mathcal{D}_{b}\omega
\nonumber
\\
&=\mathcal{D}^{c}\left(R^{(0)}_{bpca}\omega^{p}\right)+\mathcal{D}_{c}\, \mathcal{D}_{a}\, \mathcal{D}_{b}\omega^{c} 
\nonumber
\\
&=\left(-h^{(0)}_{bc}h^{(0)}_{pa}+h^{(0)}_{ba}h^{(0)}_{pc}\right)\omega^{pc}+[\mathcal{D}_{c},\mathcal{D}_{a}]\, \mathcal{D}_{b}\omega^{c}+ \mathcal{D}_{a}\, \mathcal{D}_{c} \, \mathcal{D}_{b}\omega^{c}
\nonumber
\\
&=-\omega_{ab}+h^{(0)}_{ab}\square \omega+R^{(0)}_{bpca}\omega^{cp}+R^{(0)}_{pa}\omega^{p}_{b}+ \mathcal{D}_{a}\, [\mathcal{D}_{c},\mathcal{D}_{b}]\omega^{c}+ \mathcal{D}_{a}\, \mathcal{D}_{b}\, \square \omega
\nonumber
\\
&=-\omega_{ab}+3\omega h^{(0)}_{ab}+\left(-h^{(0)}_{bc}h^{(0)}_{pa}+h^{(0)}_{ba}h^{(0)}_{pc}\right)\omega^{cp}-2\omega_{ab}+R^{(0)}_{pb}\omega^{p}_{a}+3\omega_{ab}
\nonumber
\\
&=-3\omega_{ab}+6\omega h^{(0)}_{ab}~.
\end{align}
Thus we obtain,
\begin{align}
\left(\square+2\right)h^{(2)}_{ab}=~& 8\omega^{2}h^{(0)}_{ab}-10\omega \omega_{ab}+2h^{(0)}_{ab}\left(\omega_{c}\omega^{c}\right)+2\omega^{c}_{a}\omega_{cb}-4\omega^{c}\omega_{abc}+2\omega_{acd}\omega_{b}^{\,\,cd}
\nonumber
\\
& +\omega_{cb}\left(-3\omega^{c}_{a}+6\omega \delta^{c}_{a}\right)+\omega^{c}_{a}\left(-3\omega_{cb}+6\omega h^{(0)}_{cb} \right)-2\omega\left(-3\omega_{ab}+6\omega h^{(0)}_{ab}\right)
\nonumber
\\
=~& -4\omega^{2}h^{(0)}_{ab}+8\omega \omega_{ab}+2h^{(0)}_{ab}\left(\omega_{c}\omega^{c}\right)-4\omega^{c}_{a}\omega_{cb}-4\omega^{c}\omega_{abc}+2\omega_{acd}\omega_{b}^{\,\,cd}~,
\end{align}
which matches with \ref{evo_second_super}.   A similar calculation at spatial infinity was done in \cite{Virmani:2011aa}.

 \section{Expansion of the Weyl tensor}
\label{Weyl_tensor_expansion}

In   four spacetime dimensions, the Weyl tensor expressed in terms of the Riemann tensor, Ricci tensor and Ricci scalar takes the form, 
\begin{align}\label{Weyl_Decomp}
W_{\alpha \beta \mu \nu}=R_{\alpha \beta \mu \nu}-\frac{1}{2}\left(g_{\alpha \mu}R_{\beta \nu}+R_{\alpha \mu}g_{\beta \nu}-g_{\alpha \nu}R_{\beta \mu}-R_{\alpha \nu}g_{\beta \mu}\right)+\frac{R}{6}\left(g_{\alpha \mu}g_{\beta \nu}-g_{\alpha \nu}g_{\beta \mu} \right)~.
\end{align}
 Let  $(\tau, \phi^a)$ be the four-dimensional spacetime coordinates associated to the 3+1 split. Then, for a general  set of spacetime coordinates $x^\mu= x^\mu(\tau, \phi^a)$ we define
\be
e^{\mu}_{a} = \frac{\partial x^\mu}{\partial \phi^a}.
\ee
The vectors  $e^{\mu}_{a}$ with $\{a=1,2,3\}$ are tangent to the $\tau=\textrm{constant}$ hypersurface.  
The projected electric part of the Weyl tensor on $\tau=\textrm{constant}$ hypersurface is defined as,
\begin{align}
E_{ab}&=W_{\alpha \beta \mu \nu}e^{\alpha}_{a}n^{\beta}e^{\mu}_{b}n^{\nu}.
\end{align}
For vacuum spacetimes, with $R_{\alpha \beta}=0=R$, it simplifies to,
\begin{align}
E_{ab}=R_{\alpha \beta \mu \nu}e^{\alpha}_{a}n^{\beta}e^{\mu}_{b}n^{\nu}=-\pounds_{n}K_{ab}+K_{ac}K^{c}_{b}+N^{-1}D_{a}D_{b}N~,
\end{align}
where $\pounds_{n}$ is the Lie-derivative with respect to the unit normal \ref{unit_normal}. 
We have used the fact that $\tau=\textrm{constant}$ surface is spacelike. 

The projected magnetic part of the Weyl tensor is defined as,
\begin{align}
B_{ab}&=\frac{1}{2}\left(\epsilon_{\alpha \beta \rho \sigma}W^{\rho \sigma}_{~~~~~\mu \nu}\right)e^{\alpha}_{a}n^{\beta}e^{\mu}_{b}n^{\nu}
\nonumber
\\
&=\frac{1}{2}\left(e^{\alpha}_{a}n^{\beta}\epsilon_{\alpha \beta \rho \sigma}\right)g^{\rho \gamma}g^{\sigma \delta}W_{\gamma \delta  \mu \nu}e^{\mu}_{b}n^{\nu}
\nonumber
\\
&=\frac{1}{2}\left(e^{\alpha}_{a}n^{\beta}\epsilon_{\alpha \beta \rho \sigma}\right)\left(h^{\rho \gamma}
+\epsilon n^{\rho}n^{\gamma} \right)\left(h^{\sigma \delta}+\epsilon n^{\sigma}n^{\delta}\right)W_{\gamma \delta  \mu \nu}e^{\mu}_{b}n^{\nu}
\nonumber
\\
&=\frac{1}{2}\left(e^{\alpha}_{a}n^{\beta}\epsilon_{\alpha \beta \rho \sigma}\right)\left(h^{\rho \gamma}h^{\sigma \delta}W_{\gamma \delta  \mu \nu}e^{\mu}_{b}n^{\nu}\right).
\end{align}
For vacuum spacetimes, 
\begin{align}\label{Weyl_Magnetic}
B_{ab}=\frac{1}{2}\left(e^{\alpha}_{a}n^{\beta}\epsilon_{\alpha \beta \rho \sigma}\right)\left(h^{\rho \gamma}h^{\sigma \delta}R_{\gamma \delta  \mu \nu}e^{\mu}_{b}n^{\nu}\right)=-\frac{1}{2}\epsilon _{acd}\left(D^{c}K^{d}_{b}-D^{d}K^{c}_{b}\right)=-\epsilon _{acd}D^{c}K^{d}_{b}~.
\end{align}
Note that we have used the result, $\epsilon_{\rho \alpha \beta \mu}n^{\rho}=\epsilon_{abc}e^{a}_{\alpha}e^{b}_{\beta}e^{c}_{\mu}$, where $\epsilon_{abc}$ is the three-dimensional Levi-Civita tensor. In what follows we will expand both the electric and magnetic parts of the Weyl tensor.

\subsection*{Expansion of the electric part of the Weyl tensor}

Given the expansions for the extrinsic curvature components and the lapse function $N$ in powers of $1/\tau$, we can  obtain the expansion of the electric part of the Weyl tensor. A calculation gives,
\begin{align}
E_{ab} & \equiv \frac{1}{\tau}E_{ab}^{(1)}+\frac{1}{\tau^{2}}E_{ab}^{(2)} + \cdots
\end{align}
where the zeroth order expansion coefficient identically vanishes and the first order expansion coefficient  is,
\begin{align}\label{First_electric}
E_{ab}^{(1)}=-\sigma h^{(0)}_{ab}+ \mathcal{D}_{a}\mathcal{D}_{b}\sigma,
\end{align}
while the second order expansion coefficient is,
\begin{align}
E_{ab}^{(2)} =~& 3\sigma^{2}h^{(0)}_{ab}-h^{(2)}_{ab}+\frac{1}{4}h^{(1)p}_{a}h^{(1)}_{pb} -\sigma \mathcal{D}_{a} \mathcal{D}_{b}\sigma- \Gamma^{(1)c}_{ab}\mathcal{D}_{c}\sigma-\frac{\sigma}{2}h^{(1)}_{ab}
\nonumber
\\
=~& -h^{(2)}_{ab}+5\sigma^{2}h^{(0)}_{ab}+\frac{1}{4}k^{p}_{a}k_{pb}-\sigma k_{ab}-\frac{\sigma}{2}k_{ab}-\sigma \mathcal{D}_{a} \mathcal{D}_{b}\sigma
\nonumber
\\
&  -\frac{1}{2}\left(-\mathcal{D}^{c}\, k_{ab}+ \mathcal{D}_{a}k^{c}_{b}+ \mathcal{D}_{b}k^{c}_{a}\right)\mathcal{D}_{c}\sigma+2D_{a}\sigma \mathcal{D}_{b}\sigma-h^{(0)}_{ab}\mathcal{D}_{c}\sigma \mathcal{D}^{c}\sigma.
\end{align}

For the first order term, we have the following properties, 
\begin{align}
& E^{(1)}_{ab}=E^{(1)}_{ba}, &  & (\textrm{symmetric}) &
\\
& E^{(1)a}_{a}=-3\sigma +\square \sigma =0, &  & (\textrm{traceless}) &
 \\
& \mathcal{D}_{b} E^{(1)b}_{a} = 0. & & (\textrm{divergence-free}) &
\end{align}

We consider the following combination at the second order
\begin{align}
E_{ab}^{(2)}-\sigma E_{ab}^{(1)}=~&-h^{(2)}_{ab}+6\sigma^{2}h^{(0)}_{ab}-2\sigma \mathcal{D}_{a} \mathcal{D}_{b}\sigma+2\mathcal{D}_{a}\sigma \mathcal{D}_{b} \sigma-h^{(0)}_{ab} \mathcal{D}_{c} \sigma \mathcal{D}^{c}\sigma
\nonumber
\\
&  -\frac{1}{2}\left(-\mathcal{D}^{c}k_{ab}+ \mathcal{D}_{a}k^{c}_{b}+ \mathcal{D}_{b}k^{c}_{a}\right) \mathcal{D}_{c}\sigma+\frac{1}{4}k^{p}_{a}k_{pb}-\frac{3\sigma}{2}k_{ab}
\label{second_electric_gen}
\end{align}
For $k_{ab}=0$,
\begin{align}\label{second_Electric_Weyl}
E_{ab}^{(2)}-\sigma E_{ab}^{(1)}=-h^{(2)}_{ab}+6\sigma^{2}h^{(0)}_{ab}-2\sigma \mathcal{D}_{a} \mathcal{D}_{b} \sigma+2\mathcal{D}_{a} \sigma \mathcal{D}_{b}\sigma-h^{(0)}_{ab}\mathcal{D}_{c} \sigma \mathcal{D}^{c}\sigma~,
\end{align}
 is also symmetric, traceless, and divergence free upon using second order equations of motion. The trace and divergence equations for $h^{(2)}_{ab}$ can equivalently be thought of as tracefree and divergence free conditions  for $E_{ab}^{(2)}-\sigma E_{ab}^{(1)}$.

\subsection*{Expansion of the magnetic part of the Weyl tensor}
 
We now compute the expansion of the magnetic part of the Weyl tensor starting from \ref{Weyl_Magnetic}, 
\be 
B_{ab}=-\epsilon _{acd}h^{cm}D_{m}K^{d}_{b} \equiv \frac{1}{\tau}B^{(1)}_{ab}+\frac{1}{\tau^{2}}B^{(2)}_{ab} + \cdots~.
\ee
The first order expansion coefficient is,
\begin{align}
B^{(1)}_{ab}&=\epsilon^{(0)}_{acd}\left(\frac{1}{2}\mathcal{D}^{c}h^{(1)d}_{b}+\delta^{d}_{b} \mathcal{D}^{c}\sigma \right)
\nonumber
\\
&=\epsilon^{(0)}_{acd}\left[\frac{1}{2}\mathcal{D}^{c}\left(k^{d}_{b}-2\sigma \delta^{d}_{b}\right)+\delta^{d}_{b} \mathcal{D}^{c}\sigma \right]
=\frac{1}{2}\epsilon^{(0)}_{acd}\left(\mathcal{D}^{c}k^{d}_{b}\right),
\label{first_magnetic_general}
\end{align}
while the second order expansion coefficient is, 
\begin{align}
B^{(2)}_{ab}&=\epsilon^{(0)}_{acd}\Bigg\{\left[\mathcal{D}^{c}h^{(2)d}_{b}-2\delta^{d}_{b}\mathcal{D}^{c}\left(\sigma^{2}\right) \right]
-\frac{1}{2}\left(k^{c}_{m}+\sigma \delta^{c}_{m}\right)\mathcal{D}^{m}k^{d}_{b}
+\frac{1}{2}h^{(0)cm}\Gamma^{(1)d}_{mp}k^{p}_{b}
\nonumber
\\
&\hskip 1 cm -\frac{1}{2}h^{(0)cm}\Gamma^{(1)p}_{mb}k^{d}_{p} - \mathcal{D}^{c}\left(\frac{\sigma}{2}k^{d}_{b}\right)- \mathcal{D}^{c}\left(\frac{1}{2}k^{dp}\,k_{pb}\right)+ \mathcal{D}^{c}\left(2\sigma k^{d}_{b}\right)\Bigg\}
\label{second_magnetic_gen}
\end{align}
where, we have used the result, $h^{(1)dp}h^{(1)}_{pb}=(k^{dp}-2\sigma h^{(0)dp})(k_{pb}-2\sigma h^{(0)}_{pb})=k^{dp}\,k_{pb}-4\sigma k^{d}_{b}+4\sigma^{2}\delta^{d}_{b}$. These expressions become much simpler for $k_{ab}=0$, in which case, we have, 
\begin{align}
B^{(1)}_{ab}&=0,
\label{first_magnetic}
\\
B^{(2)}_{ab}&=\epsilon^{(0)}_{acd}\,\mathcal{D}^{c}\left(h^{(2)d}_{b}-2\delta^{d}_{b}\sigma^{2}\right).
\label{second_magnetic}
\end{align}
$B^{(2)}_{ab}$ in \ref{second_magnetic} is  symmetric,
\begin{align}
\epsilon^{(0)abp}B^{(2)}_{ab}&=\epsilon^{(0)abp}\epsilon^{(0)}_{acd}\,\mathcal{D}^{c}\left(h^{(2)d}_{b}-2\delta^{d}_{b}\sigma^{2}\right) \\
&=\left(\delta^{b}_{c}\delta^{p}_{d}-\delta^{b}_{d}\delta^{p}_{c}\right)\mathcal{D}^{c}\left(h^{(2)d}_{b}-2\delta^{d}_{b}\sigma^{2}\right)
\nonumber
 \\
&=\left(\mathcal{D}^{b}h^{(2)p}_{b}\right)+4\left(\mathcal{D}^{b}\sigma^{2}\right)- \mathcal{D}^{p}\left(h^{(2)}\right)
\nonumber
\\
&= \mathcal{D}^{p}\left(8\sigma^{2}- \mathcal{D}_{c}\sigma \mathcal{D}^{c}\sigma\right)+4\left(\mathcal{D}^{b}\sigma^{2}\right)- \mathcal{D}^{p}\left(12\sigma^{2}- \mathcal{D}_{c}\sigma \mathcal{D}^{c}\sigma\right)  =0,
\end{align}
where we have used the second order equations of motion.
  $B^{(2)}_{ab}$ is  traceless,
\be
B^{(2)a}_{a}=\epsilon^{(0)}_{acd}\,\mathcal{D}^{c}\left(h^{(2)ad}-2h^{(0)ad}\sigma^{2}\right)=0,
\ee
furthermore $B^{(2)}_{ab}$ is  divergence-free,
\begin{align}
\mathcal{D}_{a}B^{(2)a}_{b}&=\epsilon^{(0)acd}\,\mathcal{D}_{a}\left[\mathcal{D}_{c}\left(h^{(2)}_{bd}-2h^{(0)}_{bd}\sigma^{2}\right)\right]
\nonumber
\\
&=\epsilon^{(0)acd}\,\mathcal{D}_{a}\left[\mathcal{D}_{c}\left(h^{(2)}_{bd}-2h^{(0)}_{bd}\sigma^{2}\right)\right]
\nonumber
\\
&=\frac{1}{2}\epsilon^{(0)acd}[\mathcal{D}_{a},\mathcal{D}_{c}]h^{(2)}_{bd}-2h^{(0)}_{bd}\epsilon^{(0)acd}\,\mathcal{D}_{a}\mathcal{D}_{c}\sigma^{2}
\nonumber
\\
&=\frac{1}{2}\epsilon^{(0)acd}\left(R_{bpac}h^{(2)p}_{d}+R_{dpac}h^{(2)p}_{b}\right)
\nonumber
\\
&=\frac{1}{2}\epsilon^{(0)acd}\left[\left(-h^{(0)}_{ab}h^{(0)}_{pc}+h^{(0)}_{bc}h^{(0)}_{pa}\right)h^{(2)p}_{d}
+\left(-h^{(0)}_{da}h^{(0)}_{pc}+h^{(0)}_{dc}h^{(0)}_{pa}\right)h^{(2)p}_{b}\right] =0.
\end{align}
The trace and divergence equations for $h^{(2)}_{ab}$ can equivalently be thought of as tracefree and divergence free conditions  for $B_{ab}^{(2)}$.


\subsection*{Evolution of the electric and magnetic parts of the Weyl tensor}

Here we describe the evolution equation for the electric and magnetic parts of the Weyl tensor. It will be advantageous to define, 
\begin{align}
\textrm{curl}~T_{ab}&=\epsilon^{(0)}_{acd}\mathcal{D}^{c}T^{d}_{b}~.
\end{align}
It follows that,
\begin{align}
\textrm{curl}\left(\textrm{curl}~T_{ab}\right)&=\epsilon^{(0)}_{acd}\,\mathcal{D}^{c}\left(\epsilon^{(0) dpq}\,\mathcal{D}_{p}T_{qb}\right)
\nonumber
\\
&=\epsilon^{(0)}_{dac}\epsilon^{(0) dpq}\,\mathcal{D}^{c}\mathcal{D}_{p}T_{qb}=\left(\delta^{p}_{a}\delta^{q}_{c}-\delta^{q}_{a}\delta^{p}_{c}\right)\mathcal{D}^{c}\mathcal{D}_{p}T_{qb}
\nonumber
\\
&=\mathcal{D}^{c}\mathcal{D}_{a}T_{cb}-\square^{(3)}T_{ab}
\nonumber
\\
&=[\mathcal{D}^{c}, \mathcal{D}_{a}]T_{cb} + \mathcal{D}_{a} \left(\mathcal{D}^{c}T_{cb}\right)-\square^{(3)}T_{ab}
\nonumber
\\
&=R^{(0)~~mc}_{~~~~c~~~~a}\,T_{mb}+R^{(0)~~mc}_{~~~~b~~~~a}\,T_{cm}+ \mathcal{D}_{a}\left(\mathcal{D}^{c}T_{cb}\right)-\square^{(3)}T_{ab}
\nonumber
\\
&=-2T_{ab}+\left(-\delta^{c}_{b}\delta^{m}_{a}+h^{(0)}_{ab}h^{(0)mc} \right)T_{cm}+ \mathcal{D}_{a} \left(\mathcal{D}^{c}T_{cb}\right)-\square^{(3)}T_{ab}
\nonumber
\\
&=-\left(\square +3\right)T_{ab} + \mathcal{D}_{a} \left(\mathcal{D}^{c}T_{cb}\right)+h^{(0)}_{ab}T^{c}_{c}~.
\end{align}
Thus, if the tensor $T_{ab}$ is traceless and divergence free, the above expression yields, 
\begin{align}\label{curl_of_curl}
\textrm{curl}\left(\textrm{curl}~T_{ab}\right)=-\left(\square +3\right)T_{ab}~.
\end{align}
Since the combination $(E^{(2)}_{ab}-\sigma E^{(1)}_{ab})$ and $B^{(2)}_{ab}$ are both traceless and divergence free, both of them satisfy the above identity.  
 A calculation then shows that
\begin{align}\label{curl_electric_final}
\textrm{curl}~\left(E_{ab}^{(2)}-\sigma E_{ab}^{(1)}\right)=-B_{ab}^{(2)}-4\epsilon_{~~~(a}^{(0)~cd}\left(\mathcal{D}_{c}\sigma \right) E^{(1)}_{b)d}~.
\end{align}
On the other hand,
\begin{align}\label{curl_magnetic}
\textrm{curl}~B_{ab}^{(2)}&=\epsilon _{~~~a}^{(0)~cd}\mathcal{D}_{c}B^{(2)}_{db}
\nonumber
\\
&=\epsilon _{~~~a}^{(0)~cd}\mathcal{D}_{c}\left[\epsilon^{(0)}_{dpq}\,\mathcal{D}^{p}\left(h^{(2)q}_{b}-2\delta^{q}_{b}\sigma^{2}\right) \right]
\nonumber
\\
&=-h^{(2)}_{ab}+6\sigma^{2}h^{(0)}_{ab}-h^{(0)}_{ab}\left(\mathcal{D}_{c}\sigma \mathcal{D}^{c}\sigma\right)+2\mathcal{D}_{a}\sigma \mathcal{D}_{b}\sigma-2\sigma \mathcal{D}_{a}\,\mathcal{D}_{b}\sigma \nonumber \\
&=E^{(2)}_{ab}-\sigma E^{(1)}_{ab}.
\end{align}
As a result, the evolution equation in terms of the electric part of the Weyl tensor takes the form
 \begin{align}\label{evolution_electric}
\Big(\square +2\Big)\left(E_{ab}^{(2)}-\sigma E_{ab}^{(1)}\right)=4~\textrm{curl}~\Big[\epsilon_{~~~(a}^{(0)~cd}\Big(\mathcal{D}_{c}\sigma \Big) E^{(1)}_{b)d} \Big]~.
\end{align}
and equivalently in terms of the magnetic part of the Weyl tensor takes the form
 \begin{align}\label{evolution_magnetic}
\left(\square +2\right)B^{(2)}_{ab}=4\epsilon^{(0)}_{cd(a} E^{(1)d}_{b)}\mathcal{D}^{c} \sigma.
\end{align}
 \ref{H2ab} can equivalently be thought of as  \ref{evolution_electric} or \ref{evolution_magnetic}. In terms of the electric and magnetic parts of the Weyl tensor, the second order equations take much simpler forms. The above analysis is inspired by the corresponding results at spacelike infinity \cite{Mann:2006bd, Mann:2008ay, Compere:2011db}.


\begin{thebibliography}{99}


\bibitem{Bondi:1962px}
H.~Bondi, M.~G.~J.~van der Burg and A.~W.~K.~Metzner,
``Gravitational waves in general relativity. 7. Waves from axisymmetric isolated systems,''
Proc. Roy. Soc. Lond. A \textbf{269}, 21-52 (1962)
doi:10.1098/rspa.1962.0161


\bibitem{Sachs:1962wk}
R.~K.~Sachs,
``Gravitational waves in general relativity. 8. Waves in asymptotically flat space-times,''
Proc. Roy. Soc. Lond. A \textbf{270}, 103-126 (1962)
doi:10.1098/rspa.1962.0206


\bibitem{Strominger:2013jfa}
A.~Strominger,
``On BMS Invariance of Gravitational Scattering,''
JHEP \textbf{07}, 152 (2014)
doi:10.1007/JHEP07(2014)152
[arXiv:1312.2229 [hep-th]].

\bibitem{He:2014laa}
T.~He, V.~Lysov, P.~Mitra and A.~Strominger,
``BMS supertranslations and Weinberg\textquoteright{}s soft graviton theorem,''
JHEP \textbf{05}, 151 (2015)
doi:10.1007/JHEP05(2015)151
[arXiv:1401.7026 [hep-th]].



\bibitem{Ashtekar:2014zsa}
A.~Ashtekar,
``Geometry and Physics of Null Infinity,''
[arXiv:1409.1800 [gr-qc]].


\bibitem{Alessio:2017lps}
F.~Alessio and G.~Esposito,
``On the structure and applications of the Bondi\textendash{}Metzner\textendash{}Sachs group,''
Int. J. Geom. Meth. Mod. Phys. \textbf{15}, no.02, 1830002 (2018)
doi:10.1142/S0219887818300027
[arXiv:1709.05134 [gr-qc]].

\bibitem{Madler:2016xju}
T.~M\"adler and J.~Winicour,
``Bondi-Sachs Formalism,''
Scholarpedia \textbf{11}, 33528 (2016)
doi:10.4249/scholarpedia.33528
[arXiv:1609.01731 [gr-qc]].

\bibitem{Strominger:2017zoo}
A.~Strominger,
``Lectures on the Infrared Structure of Gravity and Gauge Theory,''
[arXiv:1703.05448 [hep-th]].


\bibitem{Ashtekar:2018lor}
A.~Ashtekar, M.~Campiglia and A.~Laddha,
``Null infinity, the BMS group and infrared issues,''
Gen. Rel. Grav. \textbf{50}, no.11, 140-163 (2018)
doi:10.1007/s10714-018-2464-3
[arXiv:1808.07093 [gr-qc]].


\bibitem{Aneesh:2021uzk}
P.~B.~Aneesh, G.~Comp\`ere, L.~P.~de Gioia, I.~Mol and B.~Swidler,
``Celestial Holography: Lectures on Asymptotic Symmetries,''
[arXiv:2109.00997 [hep-th]].



\bibitem{Ashtekar:1981bq}
A.~Ashtekar and M.~Streubel,
``Symplectic Geometry of Radiative Modes and Conserved Quantities at Null Infinity,''
Proc. Roy. Soc. Lond. A \textbf{376}, 585-607 (1981)
doi:10.1098/rspa.1981.0109

\bibitem{Ashtekar:1981sf}
A.~Ashtekar,
``Asymptotic Quantization of the Gravitational Field,''
Phys. Rev. Lett. \textbf{46}, 573-576 (1981)
doi:10.1103/PhysRevLett.46.573

\bibitem{Ashtekar:1987tt}
A.~Ashtekar,
``ASYMPTOTIC QUANTIZATION: BASED ON 1984 NAPLES LECTURES,'' Naples, Italy: Bibliopolis (1987), Monographs and Textbooks in Physical Sciences, Lecture notes. 

\bibitem{Barnich:2010eb}
G.~Barnich and C.~Troessaert,
``Aspects of the BMS/CFT correspondence,''
JHEP \textbf{05}, 062 (2010)
doi:10.1007/JHEP05(2010)062
[arXiv:1001.1541 [hep-th]].


\bibitem{Barnich:2010ojg}
G.~Barnich and C.~Troessaert,
``Supertranslations call for superrotations,''
PoS \textbf{CNCFG2010}, 010 (2010)
doi:10.22323/1.127.0010
[arXiv:1102.4632 [gr-qc]].

\bibitem{Campiglia:2014yka}
M.~Campiglia and A.~Laddha,
``Asymptotic symmetries and subleading soft graviton theorem,''
Phys. Rev. D \textbf{90}, no.12, 124028 (2014)
doi:10.1103/PhysRevD.90.124028
[arXiv:1408.2228 [hep-th]].

\bibitem{Freidel:2021fxf}
L.~Freidel, R.~Oliveri, D.~Pranzetti and S.~Speziale,
``The Weyl BMS group and Einstein\textquoteright{}s equations,''
JHEP \textbf{07}, 170 (2021)
doi:10.1007/JHEP07(2021)170
[arXiv:2104.05793 [hep-th]].


\bibitem{Gupta:2021cwo}
N.~Gupta, P.~Paul and N.~V.~Suryanarayana,
``An $\widehat{sl_2}$ Symmetry of ${\mathbb R}^{1,3}$ Gravity,''
[arXiv:2109.06857 [hep-th]].


  
  


  
\bibitem{Friedrich:2017cjg}
H.~Friedrich,
``Peeling or not peeling\textemdash{}is that the question?,''
Class. Quant. Grav. \textbf{35}, no.8, 083001 (2018)
doi:10.1088/1361-6382/aaafdb
[arXiv:1709.07709 [gr-qc]].


\bibitem{Arnowitt:1962hi}
R.~L.~Arnowitt, S.~Deser and C.~W.~Misner,
``The Dynamics of general relativity,''
Gen. Rel. Grav. \textbf{40}, 1997-2027 (2008)
doi:10.1007/s10714-008-0661-1
[arXiv:gr-qc/0405109 [gr-qc]].


\bibitem{Regge:1974zd}
T.~Regge and C.~Teitelboim,
``Role of Surface Integrals in the Hamiltonian Formulation of General Relativity,''
Annals Phys. \textbf{88}, 286 (1974)
doi:10.1016/0003-4916(74)90404-7



\bibitem{Ashtekar:1978zz}
  A.~Ashtekar and R.~O.~Hansen,
  ``A unified treatment of null and spatial infinity in general relativity. I - Universal structure, asymptotic symmetries, and conserved quantities at spatial infinity,''
  J.\ Math.\ Phys.\  {\bf 19} (1978) 1542.
  
\bibitem{Ashtekar:1991vb}
  A.~Ashtekar and J.~D.~Romano,
  ``Spatial infinity as a boundary of space-time,''
  Class.\ Quant.\ Grav.\  {\bf 9} (1992) 1069.



\bibitem{Troessaert:2017jcm}
  C.~Troessaert,
  ``The BMS4 algebra at spatial infinity,''
  Class.\ Quant.\ Grav.\  {\bf 35} (2018) no.7,  074003
   [arXiv:1704.06223 [hep-th]].
  
  
\bibitem{Henneaux:2018cst}
  M.~Henneaux and C.~Troessaert,
  ``BMS Group at Spatial Infinity: the Hamiltonian (ADM) approach,''
  JHEP {\bf 1803} (2018) 147
  [arXiv:1801.03718 [gr-qc]].
    
   

\bibitem{Henneaux:2018hdj}
M.~Henneaux and C.~Troessaert,
``Hamiltonian structure and asymptotic symmetries of the Einstein-Maxwell system at spatial infinity,''
JHEP \textbf{07}, 171 (2018)
doi:10.1007/JHEP07(2018)171
[arXiv:1805.11288 [gr-qc]].

  
  
  
\bibitem{Prabhu:2019fsp}
K.~Prabhu,
``Conservation of asymptotic charges from past to future null infinity: Supermomentum in general relativity,''
JHEP \textbf{03}, 148 (2019)
doi:10.1007/JHEP03(2019)148
[arXiv:1902.08200 [gr-qc]].



\bibitem{Prabhu:2019daz}
K.~Prabhu and I.~Shehzad,
``Asymptotic symmetries and charges at spatial infinity in general relativity,''
Class. Quant. Grav. \textbf{37}, no.16, 165008 (2020)
doi:10.1088/1361-6382/ab954a
[arXiv:1912.04305 [gr-qc]].
 
K.~Prabhu and I.~Shehzad,
``Conservation of asymptotic charges from past to future null infinity: Lorentz charges in general relativity,''
[arXiv:2110.04900 [gr-qc]].


\bibitem{cutler}
C.~Cutler, ``Properties of spacetimes that are asymptotically flat at timelike infinity,'' Class.~Quant.~Grav. {\bf 6} 1075.


\bibitem{porrill}
J.~Porrill, ``The structure of timelike infinity for isolated systems,'' Proc. R. Soc. Lond. A {\bf 381} 323--344.

  
\bibitem{Gen:1997az}
U.~Gen and T.~Shiromizu,
``Timelike infinity and asymptotic symmetry,''
J. Math. Phys. \textbf{39}, 6573-6592 (1998)
doi:10.1063/1.532666
[arXiv:gr-qc/9709009 [gr-qc]].





\bibitem{Campiglia:2015kxa}
M.~Campiglia and A.~Laddha,
``Asymptotic symmetries of gravity and soft theorems for massive particles,''
JHEP \textbf{12}, 094 (2015)
doi:10.1007/JHEP12(2015)094
[arXiv:1509.01406 [hep-th]].


\bibitem{H:2020eei}
A.~A.H., A.~Khairnar and A.~Kundu,
``Generalized BMS algebra at timelike infinity,''
Phys. Rev. D \textbf{103}, no.10, 104030 (2021)
doi:10.1103/PhysRevD.103.104030
[arXiv:2005.05209 [hep-th]].



\bibitem{Koga:2001vq} 
  J.~i.~Koga,
  ``Asymptotic symmetries on Killing horizons,''
  Phys.\ Rev.\ D {\bf 64}, 124012 (2001)
  doi:10.1103/PhysRevD.64.124012
  [gr-qc/0107096].



\bibitem{Donnay:2015abr} 
  L.~Donnay, G.~Giribet, H.~A.~Gonzalez and M.~Pino,
  ``Supertranslations and Superrotations at the Black Hole Horizon,''
  Phys.\ Rev.\ Lett.\  {\bf 116}, no. 9, 091101 (2016)
  doi:10.1103/PhysRevLett.116.091101
  [arXiv:1511.08687 [hep-th]].




\bibitem{Hawking:2016msc} 
  S.~W.~Hawking, M.~J.~Perry and A.~Strominger,
  ``Soft Hair on Black Holes,''
  Phys.\ Rev.\ Lett.\  {\bf 116}, no. 23, 231301 (2016)
  doi:10.1103/PhysRevLett.116.231301
  [arXiv:1601.00921 [hep-th]].


\bibitem{Hawking:2016sgy} 
  S.~W.~Hawking, M.~J.~Perry and A.~Strominger,
  ``Superrotation Charge and Supertranslation Hair on Black Holes,''
  JHEP {\bf 1705}, 161 (2017)
  doi:10.1007/JHEP05(2017)161
  [arXiv:1611.09175 [hep-th]].


\bibitem{Carlip:2017xne} 
  S.~Carlip,
 ``Black Hole Entropy from Bondi-Metzner-Sachs Symmetry at the Horizon,''
  Phys.\ Rev.\ Lett.\  {\bf 120}, no. 10, 101301 (2018)
  doi:10.1103/PhysRevLett.120.101301
  [arXiv:1702.04439 [gr-qc]].

   

\bibitem{Chandrasekaran:2018aop} 
  V.~Chandrasekaran, \'E.~\'E.~Flanagan and K.~Prabhu,
  ``Symmetries and charges of general relativity at null boundaries,''
  JHEP {\bf 1811}, 125 (2018)
  doi:10.1007/JHEP11(2018)125
  [arXiv:1807.11499 [hep-th]].





\bibitem{Fernandes:2020jto}
K.~Fernandes, D.~Ghosh and A.~Virmani,
``Horizon Hair from Inversion Symmetry,''
Class. Quant. Grav. \textbf{38}, no.5, 055006 (2020)
doi:10.1088/1361-6382/abd225
[arXiv:2008.04365 [hep-th]].



\bibitem{Donnay:2020fof}
L.~Donnay, G.~Giribet and F.~Rosso,
``Quantum BMS transformations in conformally flat space-times and holography,''
JHEP \textbf{12}, 102 (2020)
doi:10.1007/JHEP12(2020)102
[arXiv:2008.05483 [hep-th]].
  




\bibitem{Parattu:2015gga}
K.~Parattu, S.~Chakraborty, B.~R.~Majhi and T.~Padmanabhan,
``A Boundary Term for the Gravitational Action with Null Boundaries,''
Gen. Rel. Grav. \textbf{48}, no.7, 94 (2016)
doi:10.1007/s10714-016-2093-7
[arXiv:1501.01053 [gr-qc]].



\bibitem{Parattu:2016trq}
K.~Parattu, S.~Chakraborty and T.~Padmanabhan,
``Variational Principle for Gravity with Null and Non-null boundaries: A Unified Boundary Counter-term,''
Eur. Phys. J. C \textbf{76}, no.3, 129 (2016)
doi:10.1140/epjc/s10052-016-3979-y
[arXiv:1602.07546 [gr-qc]].





\bibitem{Hopfmuller:2016scf}
F.~Hopfm\"uller and L.~Freidel,
``Gravity Degrees of Freedom on a Null Surface,''
Phys. Rev. D \textbf{95}, no.10, 104006 (2017)
doi:10.1103/PhysRevD.95.104006
[arXiv:1611.03096 [gr-qc]].



\bibitem{Grumiller:2020vvv}
D.~Grumiller, M.~M.~Sheikh-Jabbari and C.~Zwikel,
``Horizons 2020,''
Int. J. Mod. Phys. D \textbf{29}, no.14, 2043006 (2020)
doi:10.1142/S0218271820430063
[arXiv:2005.06936 [hep-th]].



\bibitem{Adami:2021nnf}
H.~Adami, D.~Grumiller, M.~M.~Sheikh-Jabbari, V.~Taghiloo, H.~Yavartanoo and C.~Zwikel,
``Null boundary phase space: slicings, news \& memory,''
JHEP \textbf{11}, 155 (2021)
doi:10.1007/JHEP11(2021)155
[arXiv:2110.04218 [hep-th]].
  
  
     \bibitem{BeigSchmidt}
  R.~Beig and B.~Schmidt, ``Einstein's equations near spatial infinity,'' Commun. Math. Phys. {\bf 87} (1982) 65.
  
\bibitem{Beig:1983sw}
  R.~Beig,
  ``Integration Of Einstein's Equations Near Spatial Infinity,''
  Proc.  Royal Soc. A
{\bf 1801} (1984) 295--304.

     \bibitem{Ashtekar:1990gc} 
  A.~Ashtekar, L.~Bombelli and O.~Reula, ``The Covariant Phase Space Of Asymptotically Flat Gravitational Fields,''  in Analysis, Geometry and Mechanics: 200 Years After Lagrange, edited by M. Francaviglia and D. Holm (North-Holland, Amsterdam, 1991).





\bibitem{Ashtekar:2008jw}
A.~Ashtekar, J.~Engle and D.~Sloan,
``Asymptotics and Hamiltonians in a First order formalism,''
Class. Quant. Grav. \textbf{25}, 095020 (2008)
doi:10.1088/0264-9381/25/9/095020
[arXiv:0802.2527 [gr-qc]].


\bibitem{Compere:2011db}
G.~Compere, F.~Dehouck and A.~Virmani,
``On Asymptotic Flatness and Lorentz Charges,''
Class. Quant. Grav. \textbf{28}, 145007 (2011)
doi:10.1088/0264-9381/28/14/145007
[arXiv:1103.4078 [gr-qc]].


\bibitem{Compere:2011ve}
G.~Compere and F.~Dehouck,
``Relaxing the Parity Conditions of Asymptotically Flat Gravity,''
Class. Quant. Grav. \textbf{28}, 245016 (2011)
[erratum: Class. Quant. Grav. \textbf{30}, 039501 (2013)]
doi:10.1088/0264-9381/28/24/245016
[arXiv:1106.4045 [hep-th]].




\bibitem{Virmani:2011aa}
A.~Virmani,
``Supertranslations and Holographic Stress Tensor,''
JHEP \textbf{02}, 024 (2012)
doi:10.1007/JHEP02(2012)024
[arXiv:1112.2146 [hep-th]].


\bibitem{Campiglia:2015lxa}
M.~Campiglia,
``Null to time-like infinity Green\textquoteright{}s functions for asymptotic symmetries in Minkowski spacetime,''
JHEP \textbf{11}, 160 (2015)
doi:10.1007/JHEP11(2015)160
[arXiv:1509.01408 [hep-th]].




\bibitem{Mann:2006bd}
R.~B.~Mann, D.~Marolf and A.~Virmani,
``Covariant Counterterms and Conserved Charges in Asymptotically Flat Spacetimes,''
Class. Quant. Grav. \textbf{23}, 6357-6378 (2006)
doi:10.1088/0264-9381/23/22/017
[arXiv:gr-qc/0607041 [gr-qc]].



\bibitem{Mann:2008ay}
R.~B.~Mann, D.~Marolf, R.~McNees and A.~Virmani,
``On the Stress Tensor for Asymptotically Flat Gravity,''
Class. Quant. Grav. \textbf{25}, 225019 (2008)
doi:10.1088/0264-9381/25/22/225019
[arXiv:0804.2079 [hep-th]].


\bibitem{Lee:1990nz} 
  J.~Lee and R.~M.~Wald, ``Local symmetries and constraints,''
  J.\ Math.\ Phys.\  {\bf 31}, 725 (1990).
  doi:10.1063/1.528801
  
  
\bibitem{Iyer:1994ys} 
  V.~Iyer and R.~M.~Wald,
  ``Some properties of Noether charge and a proposal for dynamical black hole entropy,''
  Phys.\ Rev.\ D {\bf 50}, 846 (1994)
  doi:10.1103/PhysRevD.50.846
  [gr-qc/9403028].



\bibitem{Compere:2018aar}
G.~Comp\`ere and A.~Fiorucci,
``Advanced Lectures on General Relativity,''
[arXiv:1801.07064 [hep-th]].


 
\end{thebibliography}
\end{document}